\documentclass[article,
 amsmath,amssymb,
 aps,
]{revtex4-2}

\usepackage{graphicx}
\usepackage{dcolumn}
\usepackage{bm}
\usepackage{xcolor}

\begin{document}


\title{A universal angular-dispersion synthesizer}

\author{Layton A. Hall$^{1}$}
\author{Ayman F. Abouraddy$^{1,*}$}
\affiliation{$^{1}$CREOL, The College of Optics \& Photonics, University of Central~Florida, Orlando, FL 32816, USA}
\affiliation{$^*$Corresponding author: raddy@creol.ucf.edu}

\begin{abstract}
We uncover a surprising gap in optics with regards to angular dispersion (AD) that has persisted for decades. A systematic examination of pulsed optical-field configurations classified according to their three lowest dispersion orders resulting from AD (the axial phase velocity, group velocity, and group-velocity dispersion) reveals that the majority of possible classes of fields have eluded optics thus far. This gap is due in part to the limited technical reach of the standard components that provide AD such as gratings and prisms, but due in part also to misconceptions regarding the set of physically admissible field configurations that can be accessed via AD. For example, it has long been thought that AD cannot yield normal group-velocity dispersion in free space. To rectify this state of affairs, we introduce a `universal AD synthesizer': a pulsed-beam shaper that produces a wavelength-dependent propagation angle with arbitrary spectral profile, thereby enabling access to all physically admissible field configurations realizable via AD. This universal AD synthesizer is a versatile tool for preparing pulsed optical fields for dispersion compensation, optical signal processing, and nonlinear optics.
\end{abstract}


\maketitle

\section{Introduction}

Angular dispersion (AD) has remained a ubiquitous optical effect since its inception by Newton in the course of his prism experiments \cite{Sabra81Book}. By AD we refer to the wavelength-dependence of the propagation angle in polychromatic fields, which is introduced by diffractive or dispersive components such as gratings or prisms \cite{Fulop10Review,Torres10AOP}. In general, AD can help change the group velocity \cite{Hebling02OE} or produce group-velocity dispersion (GVD) \cite{Szatmari96OL}, leading to a wide range of applications in dispersion compensation \cite{Martinez84JOSAA,Fork84OL,Gordon84OL}, pulse compression \cite{Bor85OC,Lemoff93OL,Kane97JOSAB}, broadband phase-matching in nonlinear optics \cite{Martinez89IEEE,Szabo90APB,Szabo94APB,Richman98OL,Richman99AO}, and the generation of THz pulses \cite{Hebling02OE,Nugraha19OL,Wang20LPR}.

Surprisingly, despite the passage of centuries, the methodology utilized today to produce AD with prisms and gratings does not differ fundamentally from that implemented by Newton or Fraunhofer. More recently, metasurfaces have been fabricated to control the sign of the first-order AD \cite{Arbabi17Optica,McClung20Light}, or to combine the roles of a grating and a lens and produce a tilted-pulse front (TPF) for potential applications in beam steering \cite{Shaltout19ScienceSteering}. In all these cases -- whether traditional devices or metasurfaces -- only first-order AD is manipulated, but not the higher-order terms \cite{Porras03PRE2}.

This state of affairs has led to a curious gap in optics that has survived unnoticed till today. This gap can be appreciated by classifying pulsed optical fields in free space according to their first three dispersion orders engendered by AD: the axial phase velocity $v_{\mathrm{ph}}$, group velocity $\widetilde{v}$, and GVD. With respect to the axial phase velocity $v_{\mathrm{ph}}$, we divide fields into on-axis ($v_{\mathrm{ph}}\!=\!c$) and off-axis ($v_{\mathrm{ph}}\!\neq\!c$) classes, where $c$ is the speed of light in vacuum; with respect to the axial group velocity $\widetilde{v}$, we have luminal ($\widetilde{v}\!=\!c$) and non-luminal ($\widetilde{v}\!\neq\!c$) classes; and with respect to dispersion, we have fields that are dispersion-free, endowed with anomalous or normal GVD, or have an arbitrary dispersion profile. According to this scheme, optical fields endowed with AD fall into $2\times2\times4\!=\!16$ possible classes. Surprisingly, we find that \textit{the majority of classes of physically admissible field configurations from this classification have yet to be realized}. In fact, representatives from only 6 classes have been synthesized to date, and of the remaining 10 classes only one is physically excluded -- the other 9 classes have thus far eluded optics.

Two factors have contributed to this surprising situation. First, misconceptions have persisted for decades with regards to the set of physically admissible optical fields that can be realized via AD. Specifically, the result by Martinez, Gordon, and Fork \cite{Martinez84JOSAA} purports to show that AD yields \textit{only} anomalous GVD in free space, a result that forms the basis for the utilization of prism pairs \cite{Fork84OL} and other optical systems \cite{Gordon84OL} to compensate for normal material GVD. Second, conventional optical components such as gratings and prisms offer limited control over the spectral profile of AD.

Producing all the physically realizable field configurations accessible via AD requires: (1) independent tuning of multiple orders of AD, a feature that is not provided by conventional optical components; and (2) access to the newly identified \textit{non-differentiable} AD, whereby the derivative of the wavelength-dependent propagation angle is undefined at some wavelength \cite{Hall21OL,Yessenov21ACSP,Hall21OL3NormalGVD,Hall21OE1NonDiff}, a condition that is not produced by any currently available optical device. Non-differentiable AD arises naturally in the study of `space-time' (ST) wave packets \cite{Kondakci16OE,Parker16OE,Kondakci17NP,Porras17OL,Efremidis17OL,Yessenov19OPN,Wong21OE}, where it undergirds their unique characteristics in free space such as propagation invariance \cite{Kondakci18PRL,Bhaduri19OL,Yessenov19OE,Yessenov19Optica,Yessenov20NC,Schepler20ACSP,Wong20AS}, tunable group velocity \cite{Wong17ACSP2,Porras17OL,Efremidis17OL,Kondakci19NC}, self-healing \cite{Kondakci18OL}, Talbot self-imaging in space-time \cite{Hall21APLP,Hall21OL2}, accelerating wave packets \cite{Yessenov20PRL2,Hall21OL4Acceleration}, arbitrary dispersion profiles \cite{Malaguti08OL,Malaguti09PRA,Yessenov21ACSP,Hall21OL3NormalGVD}, and anomalous refraction \cite{Bhaduri20NP}. 

We show here that the versatile pulsed-beam shaper developed for the synthesis of ST wave packets \cite{Yessenov19OPN} constitutes a `universal AD synthesizer': it can produce arbitrary AD spectral profiles by controlling the relative weights of the individual AD orders. This pulsed-beam shaper comprises spectral analysis followed by wave-front phase modulation to produce differentiable or non-differentiable AD in the paraxial regime. Using this universal AD synthesizer, we produce representative wave packets from all 15 physically admissible classes of pulsed optical fields from the possible 16 classified according to their axial phase velocity, group velocity, and dispersion profile. By bridging this gap that has persisted for decades in optics, an entirely new toolbox is made available: pulsed beams with arbitrary and readily tunable dispersion characteristics. Such pulsed fields may provide new opportunities in dispersion compensation, nonlinear and quantum optics, micro-particle manipulation, light-matter interactions, and optical signal processing.

\section{Theory of angular dispersion: differentiable and non-differentiable}

We consider scalar optical fields involving one transverse spatial coordinate $x$, while holding the field uniform along $y$, and $z$ is the axial coordinate [Fig.~\ref{Fig:FieldConfiguration}]. If each temporal frequency $\omega$ in presence of AD travels at an angle $\varphi(\omega)$ with respect to the $z$-axis, then the field is $E(x,z;t)\!=\!\int\!d\omega\widetilde{E}(\omega)e^{ik(x\sin\{\varphi(\omega)\}+z\cos\{\varphi(\omega)\}-ct)}$, where $\widetilde{E}(\omega)$ is the Fourier transform of $E(0,0;t)$, $k\!=\!\omega/c$, and the transverse and longitudinal components of the wave vector are $k_{x}(\omega)\!=\!k\sin\{\varphi(\omega)\}$ and $k_{z}(\omega)\!=\!k\cos\{\varphi(\omega)\}$, respectively. We expand $\varphi(\omega)$ around a carrier frequency $\omega_{\mathrm{o}}$: $\varphi(\omega)\!=\!\varphi(\omega_{\mathrm{o}}+\Omega)=\varphi_{\mathrm{o}}+\varphi_{\mathrm{o}}^{(1)}\Omega+\tfrac{1}{2}\varphi_{\mathrm{o}}^{(2)}\Omega^{2}+\cdots$; where $\Omega\!=\!\omega-\omega_{\mathrm{o}}$, $\varphi_{\mathrm{o}}\!=\!\varphi(\omega_{\mathrm{o}})$, $\varphi_{\mathrm{o}}^{(n)}\!=\!\tfrac{d^{n}\varphi}{d\omega^{n}}\big|_{\omega=\omega_{\mathrm{o}}}$, and we expand $k_{x}(\omega)$ and $k_{z}(\omega)$ in terms of transverse and axial dispersion coefficients, respectively \cite{Porras03PRE2}:
\begin{equation}
k_{x}(\omega)=k_{x}^{(0)}+k_{x}^{(1)}\Omega+\tfrac{1}{2}k_{x}^{(2)}+\cdots,\;\;\;k_{z}(\omega)=k_{z}^{(0)}+k_{z}^{(1)}\Omega+\tfrac{1}{2}k_{z}^{(2)}+\cdots.
\end{equation}

\subsection{Phase velocity}

\begin{figure}[t!]
\centering
\includegraphics[width=11cm]{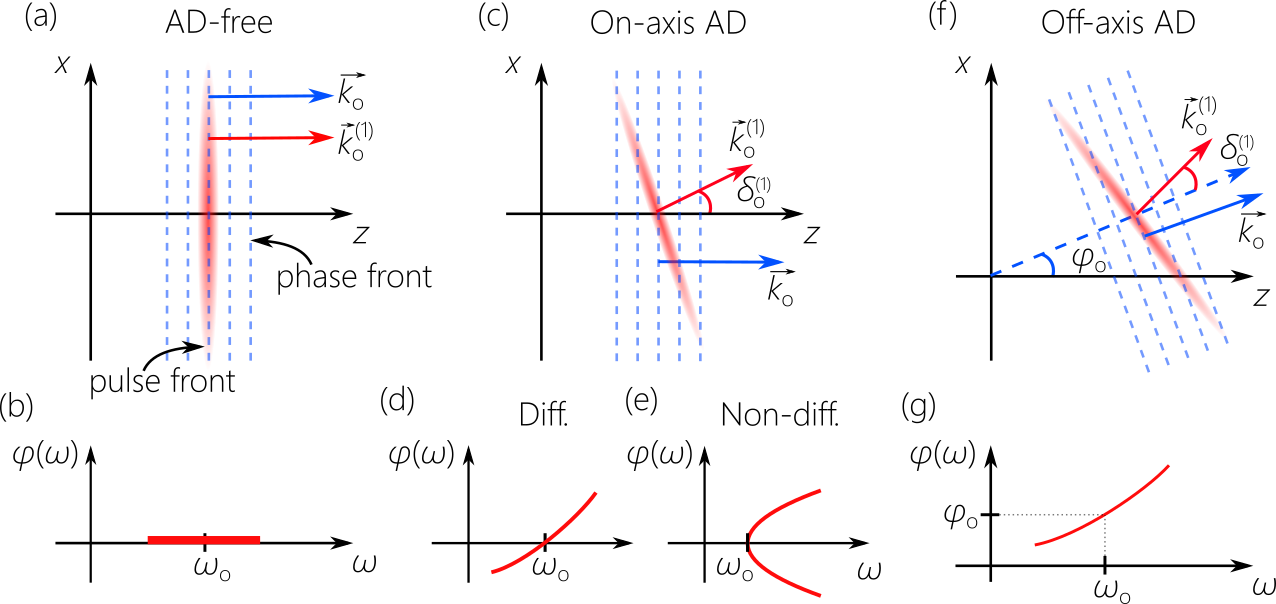}
\caption{(a) Intensity profile for a pulsed optical field free of AD, and (b) the associated propagation angle $\varphi(\omega)$. (c) On-axis ($\varphi_{\mathrm{o}}\!=\!0$) pulsed field endowed with AD that may be (d) differentiable or (e) non-differentiable. (f) Off-axis ($\varphi_{\mathrm{o}}\!\neq\!0$) pulsed field endowed with AD and (g) the associated propagation angle $\varphi(\omega)$.}
\label{Fig:FieldConfiguration}
\end{figure}

The zeroth-order dispersion terms in free space arising from AD are $k_{x}^{(0)}\!=\!k_{\mathrm{o}}\sin{\varphi_{\mathrm{o}}}$ and $k_{z}^{(0)}=k_{\mathrm{o}}\cos{\varphi_{\mathrm{o}}}$, where $k_{\mathrm{o}}\!=\!\omega_{\mathrm{o}}/c$. We refer to the condition $\varphi_{\mathrm{o}}\!=\!0$ as the `on-axis' configuration [Fig.~\ref{Fig:FieldConfiguration}(c-e)], and to $\varphi_{\mathrm{o}}\!\neq\!0$ as `off-axis' [Fig.~\ref{Fig:FieldConfiguration}(f,g)]. For large $\varphi_{\mathrm{o}}$, the field can still be useful for interacting with localized structures, but a significant propagation distance requires $\varphi_{\mathrm{o}}$ to be small. We define the vector $\vec{k}_{\mathrm{o}}\!=\!(k_{x}^{(0)},k_{z}^{(0)})$ that is orthogonal to the phase front (plane of constant phase) and makes an angle $\varphi_{\mathrm{o}}$ with the $z$-axis. The axial phase velocity $v_{\mathrm{ph}}\!=\!\tfrac{\omega_{\mathrm{o}}}{k_{z}^{(0)}}\!=\!c/\cos{\varphi_{\mathrm{o}}}$ is determined by only $\varphi_{\mathrm{o}}$ \cite{Chiao02OPN}. Therefore, $v_{\mathrm{ph}}\!=\!c$ for on-axis fields $\varphi_{\mathrm{o}}\!=\!0$; otherwise, $v_{\mathrm{ph}}\!=\!c/\cos{\varphi_{\mathrm{o}}}\!\neq\!c$ for off-axis fields. The first tier in our classification regards the axial phase velocity $v_{\mathrm{ph}}$: on-axis fields $v_{\mathrm{ph}}\!=\!c$, and off-axis fields $v_{\mathrm{ph}}\!\neq\!c$.

\subsection{Group velocity}

The first-order dispersion terms in free space arising from AD are:
\begin{equation}\label{Eq:FirstOrderDispersionTerm}
ck_{x}^{(1)}=\omega_{\mathrm{o}}\varphi_{\mathrm{o}}^{(1)}\cos{\varphi_{\mathrm{o}}}+\sin{\varphi_{\mathrm{o}}},\;\;\;ck_{z}^{(1)}=\cos{\varphi_{\mathrm{o}}}-\omega_{\mathrm{o}}\varphi_{\mathrm{o}}^{(1)}\sin{\varphi_{\mathrm{o}}},
\end{equation}
which determine the transverse walk-off and the axial group velocity $\widetilde{v}$, respectively. We make use throughout of dimensionless coefficients $c\omega_{\mathrm{o}}^{n-1}k_{x}^{(n)}$, $c\omega_{\mathrm{o}}^{n-1}k_{z}^{(n)}$, and $\omega_{\mathrm{o}}^{n}\varphi_{\mathrm{o}}^{(n)}$. The pulse front (the plane of constant amplitude) is orthogonal to the vector $\vec{k}_{\mathrm{o}}^{(1)}\!=\!(k_{x}^{(1)},k_{z}^{(1)})$, which makes an angle $\delta_{\mathrm{o}}^{(1)}$ with $\vec{k}_{\mathrm{o}}$, where $\tan{\delta_{\mathrm{o}}^{(1)}}\!=\!\omega_{\mathrm{o}}\varphi_{\mathrm{o}}^{(1)}$ \cite{Hebling96OQE}. The axial group velocity is:
\begin{equation}\label{Eq:GeneralGroupVelocity}
\widetilde{v}=\frac{1}{k_{z}^{(1)}}=\frac{c}{\cos{\varphi_{\mathrm{o}}}-\omega_{\mathrm{o}}\varphi_{\mathrm{o}}^{(1)}\sin{\varphi_{\mathrm{o}}}}=\frac{\cos{\delta^{(1)}}}{\cos{(\varphi_{\mathrm{o}}+\delta_{\mathrm{o}}^{(1)})}}.
\end{equation}
Unlike $v_{\mathrm{ph}}$ that depends solely on the \textit{geometric} factor $\varphi_{\mathrm{o}}$, $\widetilde{v}$ also incorporates an \textit{interferometric} contribution $\omega_{\mathrm{o}}\varphi_{\mathrm{o}}^{(1)}$, and thus can take on luminal \textit{or} non-luminal values in both on-axis \textit{and} off-axis fields. For off-axis fields $\varphi_{\mathrm{o}}\!\neq\!0$, we have in general $\widetilde{v}\!\neq\!c$, but the luminal condition $\widetilde{v}\!=\!c$ is achieved whenever $\varphi_{\mathrm{o}}\!=\!-2\delta^{(1)}$.

At first it appears, however, that \textit{only} luminal group velocities $\widetilde{v}\!=\!c$ can be realized in on-axis fields $\varphi_{\mathrm{o}}\!=\!0$. This was the accepted wisdom until our recent development of `baseband' ST wave packets \cite{Kondakci17NP,Kondakci19NC,Yessenov19OE,Yessenov19PRA}, which are on-axis fields with tunable group velocity $\widetilde{v}\!\neq\!c$ that seem to contradict Eq.~\ref{Eq:GeneralGroupVelocity}. However, the AD underlying baseband ST wave packets is \textit{non-differentiable} at $\omega_{\mathrm{o}}$; that is, $\tfrac{d\varphi}{d\omega}$ is not defined at $\omega\!=\!\omega_{\mathrm{o}}$. Specifically, $\varphi(\omega)\!\approx\!\eta\sqrt{\tfrac{\Omega}{\omega_{\mathrm{o}}}}$ in the vicinity of $\omega_{\mathrm{o}}$, where $\eta$ is a dimensionless constant. Because $\varphi(\omega)\!\propto\!\sqrt{\Omega}$, it is not differentiable at $\omega\!=\!\omega_{\mathrm{o}}$. Nevertheless, $\varphi\tfrac{d\varphi}{d\omega}\!\rightarrow\!\tfrac{\eta^{2}}{2\omega_{\mathrm{o}}}$ is finite and frequency-independent when $\omega\!\rightarrow\!\omega_{\mathrm{o}}$, and the on-axis field is therefore no longer luminal $\widetilde{v}\!=\!c/\widetilde{n}\!\neq\!c$, with an effective group index is $\widetilde{n}\!=\!1-\tfrac{1}{2}\eta^{2}$. Because $\widetilde{v}$ here is frequency-independent, all higher-order dispersion terms are eliminated and the ST wave packet is propagation invariant. We have therefore shown that both luminal \textit{and} non-luminal on-axis fields are indeed feasible in contrast to traditional expectations.

The second tier in our classification concerns $\widetilde{v}$: we distinguish between luminal $\widetilde{v}\!=\!c$ and non-luminal $\widetilde{v}\!\neq\!c$ fields. By combining luminal or non-luminal $v_{\mathrm{ph}}$ and $\widetilde{v}$ as distinguishing criteria, optical fields can be divided into $2\times2\!=\!4$ broad distinct categories.

\subsection{Group-velocity dispersion}

The second-order dispersion terms in free space arising from AD are \cite{Porras03PRE2}:
\begin{eqnarray}\label{Eq:SecondOrderDispersionTerm}
c\omega_{\mathrm{o}}k_{x}^{(2)}&=&(\omega_{\mathrm{o}}^{2}\varphi_{\mathrm{o}}^{(2)}+2\omega_{\mathrm{o}}\varphi_{\mathrm{o}}^{(1)})\cos{\varphi_{\mathrm{o}}}-(\omega_{\mathrm{o}}\varphi_{\mathrm{o}}^{(1)})^{2}\sin{\varphi_{\mathrm{o}}},\nonumber\\
c\omega_{\mathrm{o}}k_{z}^{(2)}&=&-(\omega_{\mathrm{o}}\varphi_{\mathrm{o}}^{(1)})^{2}\cos{\varphi_{\mathrm{o}}}-(\omega_{\mathrm{o}}^{2}\varphi_{\mathrm{o}}^{(2)}+2\omega_{\mathrm{o}}\varphi_{\mathrm{o}}^{(1)})\sin{\varphi_{\mathrm{o}}},
\end{eqnarray}
which determine the GVD experienced by the field along the $x$ and $z$ axes, respectively, and depend on $\varphi_{\mathrm{o}}$, $\omega_{\mathrm{o}}\varphi_{\mathrm{o}}^{(1)}$, and $\omega_{\mathrm{o}}^{2}\varphi_{\mathrm{o}}^{(2)}$.

One misconception needs to be clarified regarding the possibility of producing normal GVD via AD. A result in \cite{Martinez84JOSAA} purports to show that AD in free space produces \textit{only} anomalous GVD. However, this result is \textit{not} universal and applies only to on-axis fields, $\varphi_{\mathrm{o}}\!=\!0$, whereupon $c\omega_{\mathrm{o}}k_{z}^{(2)}\!=\!-(\omega_{\mathrm{o}}\varphi_{\mathrm{o}}^{(1)})^{2}\!<\!0$. For \textit{off-axis} fields $\varphi_{\mathrm{o}}\!\neq\!0$ (which are \textit{not} dealt with in \cite{Martinez84JOSAA}), one may in principle produce normal GVD via AD by tuning the values of $\varphi_{\mathrm{o}}$, $\varphi_{\mathrm{o}}^{(1)}$, and $\varphi_{\mathrm{o}}^{(2)}$ independently. For example, Porras \textit{et al}. \cite{Porras03PRE2} propose setting $\varphi_{\mathrm{o}}^{(1)}\!=\!0$, so that $c\omega_{\mathrm{o}}k_{z}^{(2)}\!=\!-\omega_{\mathrm{o}}^{2}\varphi_{\mathrm{o}}^{(2)}\sin{\varphi_{\mathrm{o}}}$, whereupon normal GVD can be realized by controlling the signs of $\varphi_{\mathrm{o}}$ and $\varphi_{\mathrm{o}}^{(2)}$. This example exemplifies the challenge in producing normal GVD via AD: exquisite control over multiple orders of AD is required, which is \textit{not} offered by conventional optical components. Although the scheme does indeed yield normal GVD, it does not eliminate higher-order dispersion terms. This proposal was realized only very recently \cite{Hall21OL3NormalGVD}. 

The challenge remains, however, to produce \textit{on-axis} normal GVD. Consider a wave packet having $k_{z}\!=\!k_{\mathrm{o}}+\tfrac{\Omega}{\widetilde{v}}+\tfrac{1}{2}k_{2}\Omega^{2}$, which is intentionally terminated at second order in $\Omega$ to eliminate all higher-order dispersion terms. This dispersion profile is produced via AD if $\varphi(\omega)$ is given by:
\begin{equation}\label{Eq:AngleForArbitraryDispersion}
\sin{\{\varphi(\omega)\}}=\eta\sqrt{\frac{\Omega}{\omega_{\mathrm{o}}}}\;\;\frac{\omega_{\mathrm{o}}}{\omega}\;\;\sqrt{\left\{1+\frac{1+\widetilde{n}}{2}\frac{\Omega}{\omega_{\mathrm{o}}}+\frac{\sigma}{2}\left(\frac{\Omega}{\omega_{\mathrm{o}}}\right)^{2}\right\}\left\{1-\frac{\sigma}{1-\widetilde{n}}\frac{\Omega}{\omega_{\mathrm{o}}}\right\}},
\end{equation}
which is non-differentiable by virtue of the factor $\sqrt{\Omega}$; here $\sigma\!=\!\tfrac{1}{2}k_{2}\omega_{\mathrm{o}}c$. A universal AD synthesizer must therefore make available non-differentiable AD as a key ingredient to produce normal and anomalous GVD on-axis when $\widetilde{v}\!\neq\!c$, as shown recently in \cite{Yessenov21ACSP,Hall21OL3NormalGVD}.

With regards to GVD, we distinguish between \textit{four} distinct states: (1) dispersion-free fields where all the dispersion coefficients vanish, $k_{z}^{(n)}\!=\!0$ for $n\!\geq\!2$; (2) fields with \textit{anomalous} GVD $k_{z}^{(2)}\!<\!0$; (3) fields with \textit{normal} GVD $k_{z}^{(2)}\!>\!0$ -- regardless of the value of the higher-order dispersion coefficients; and (4) fields with an \textit{arbitrary} dispersion profile, in which multiple dispersion coefficients are specified simultaneously. 

\section{Classification of pulsed optical fields}

\begin{figure}[t!]
\centering
\includegraphics[width=13.2cm]{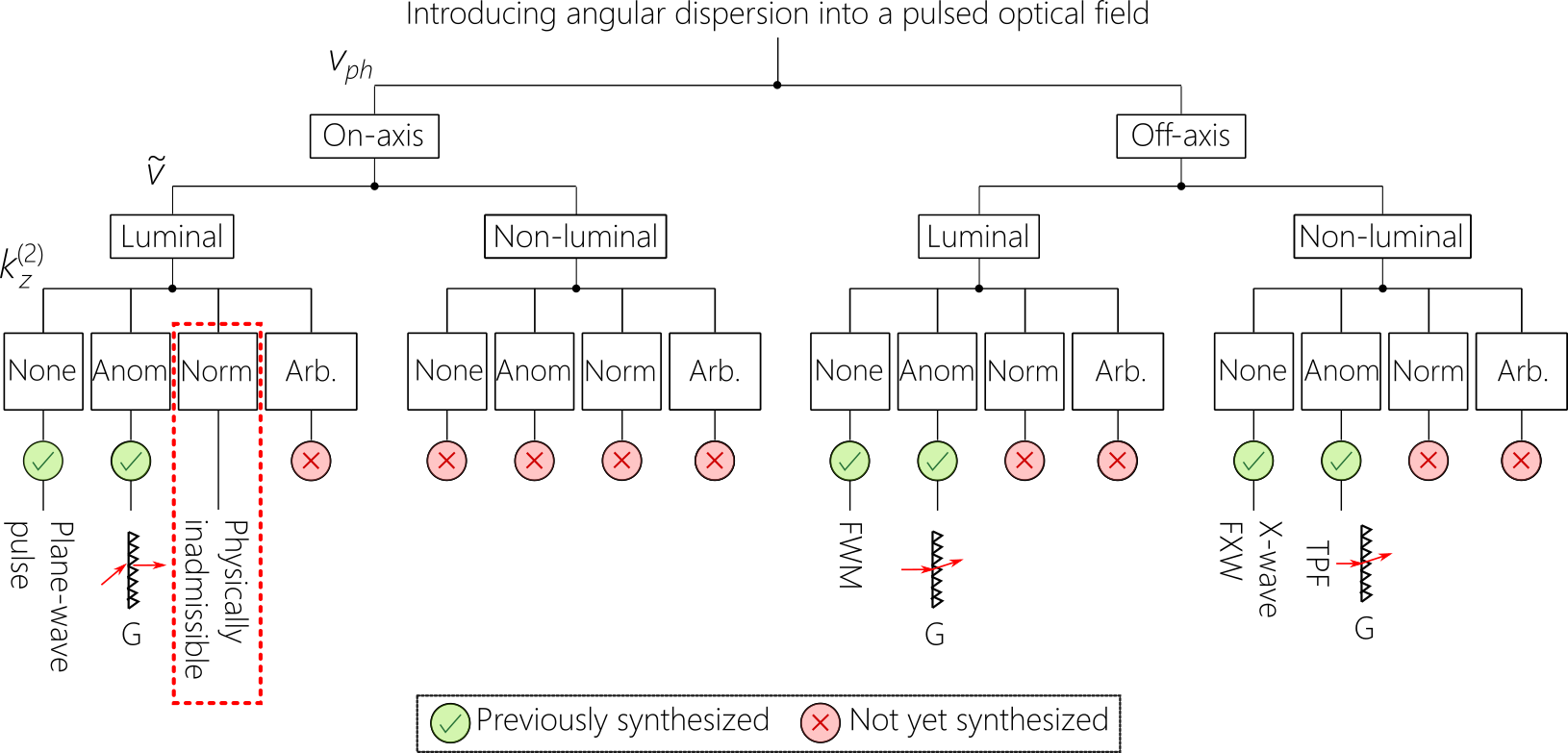}
\caption{Classification scheme for pulsed optical field configurations according to their three lowest-order dispersion terms: the axial phase velocity $v_{\mathrm{ph}}$, group velocity $\widetilde{v}$, and state of dispersion. G: Grating; FWM: focus-wave mode; FXW: focus X-wave.}
\label{Fig:Classification}
\end{figure}

We classify pulsed optical fields endowed with AD in three tiers according to the lowest dispersion orders as shown in Fig.~\ref{Fig:Classification}:
\begin{enumerate}
    \item The axial phase velocity $v_{\mathrm{ph}}\!=\!c/\cos{\varphi_{\mathrm{o}}}$: In this first tier, the fields are either on-axis ($\varphi_{\mathrm{o}}\!=\!0$ and $v_{\mathrm{ph}}\!=\!c$) or off-axis ($\varphi_{\mathrm{o}}\!\neq\!0$ and $v_{\mathrm{ph}}\!=\!c/\cos{\varphi_{\mathrm{o}}}\!\neq\!c$).
    \item The axial group velocity $\widetilde{v}$ as determined by $\varphi_{\mathrm{o}}$ and $\varphi_{\mathrm{o}}^{(1)}$ (Eq.~\ref{Eq:FirstOrderDispersionTerm}): In this second tier, the fields are luminal $\widetilde{v}\!=\!c$ or non-luminal $\widetilde{v}\!\neq\!c$.
    \item The state of axial dispersion, which we subdivide: (1) no dispersion $k_{z}^{(n)}\!=\!0$ for $n\!\geq\!2$; (2) anomalous GVD $k_{z}^{(2)}\!<\!0$, without regard to higher-order dispersion terms; (3) normal GVD $k_{z}^{(2)}\!<\!0$, without regard to higher-order dispersion terms; or (4) arbitrary dispersion profiles in which multiple dispersion orders are specified.
\end{enumerate}

According to this three-tier classification scheme, $2\times2\times4\!=\!16$ distinct classes of pulsed optical fields can be counted. The first of these 16 classes is on-axis ($\varphi_{\mathrm{o}}\!=\!0$ and $v_{\mathrm{ph}}\!=\!c$), luminal ($\widetilde{v}\!=\!c$), and dispersion-free, so that $\varphi(\omega)\!=\!0$, which corresponds to the trivial case of a plane-wave pulse traveling along the $z$-axis. Only 5 other classes of fields that are identified in Fig.~\ref{Fig:Classification} have been realized to date and have been the focus of study in the fields of AD and TPFs \cite{Fulop10Review,Torres10AOP}. Our systematic survey reveals that only one class is \textit{physically inadmissible}: on-axis ($v_{\mathrm{ph}}\!=\!c$) luminal ($\widetilde{v}\!=\!c$) fields having normal GVD, which is the particular field configuration ruled out in \cite{Martinez84JOSAA}. The 9 classes that were \textit{not} previously realized using conventional means comprise 3 classes with normal GVD and the 4 classes involving arbitrary dispersion, in addition to the dispersion-free and anomalous-GVD classes associated with on-axis non-luminal fields. These missing pulsed field configurations have either been recently realized by our group in the course of studying ST wave packets, or are reported here for the first time to the best of our knowledge. 

We proceed to examine these 16 classes of pulsed fields in terms of the projection of their spatio-temporal spectrum onto the $(k_{z},\tfrac{\omega}{c})$-plane \cite{Donnelly93ProcRSLA,Yessenov19PRA}. Because $k_{z}\!=\!\tfrac{\omega}{c}\cos{\{\varphi(\omega)\}}$, the spectral projection takes the form of a 1D curved trajectory, which must lie above the light-line $k_{z}\!=\!\tfrac{\omega}{c}$. Any point \textit{on} the light-line corresponds to $\varphi(\omega)\!=\!0$, and thus belongs to an on-axis field configuration; the spectral trajectory for off-axis fields lies away from the light-line.

\begin{figure}[t!]
\centering
\includegraphics[width=13.2cm]{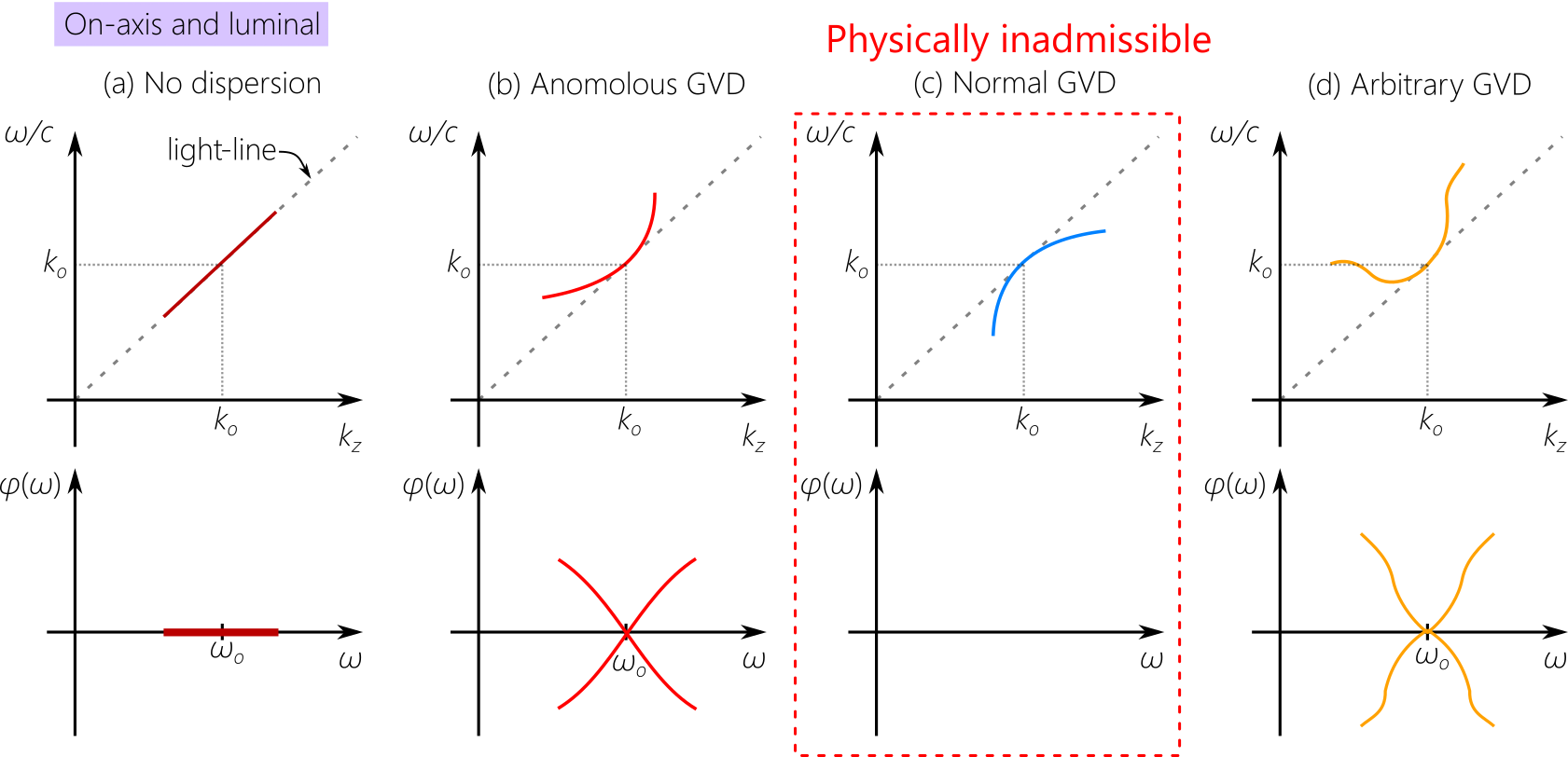}
\caption{Pulsed on-axis ($\varphi_{\mathrm{o}}\!=\!0$ and $v_{\mathrm{ph}}\!=\!c$) fields that are luminal ($\widetilde{v}\!=\!c$). In each panel we plot the spectral trajectory in the $(k_{z},\tfrac{\omega}{c})$-plane and the associated propagation angle $\varphi(\omega)$. The dashed line is the light-line $k_{z}\!=\!\tfrac{\omega}{c}$.}
\label{Fig:OnAxisLuminal}
\end{figure}

\begin{figure}[b!]
\centering
\includegraphics[width=13.2cm]{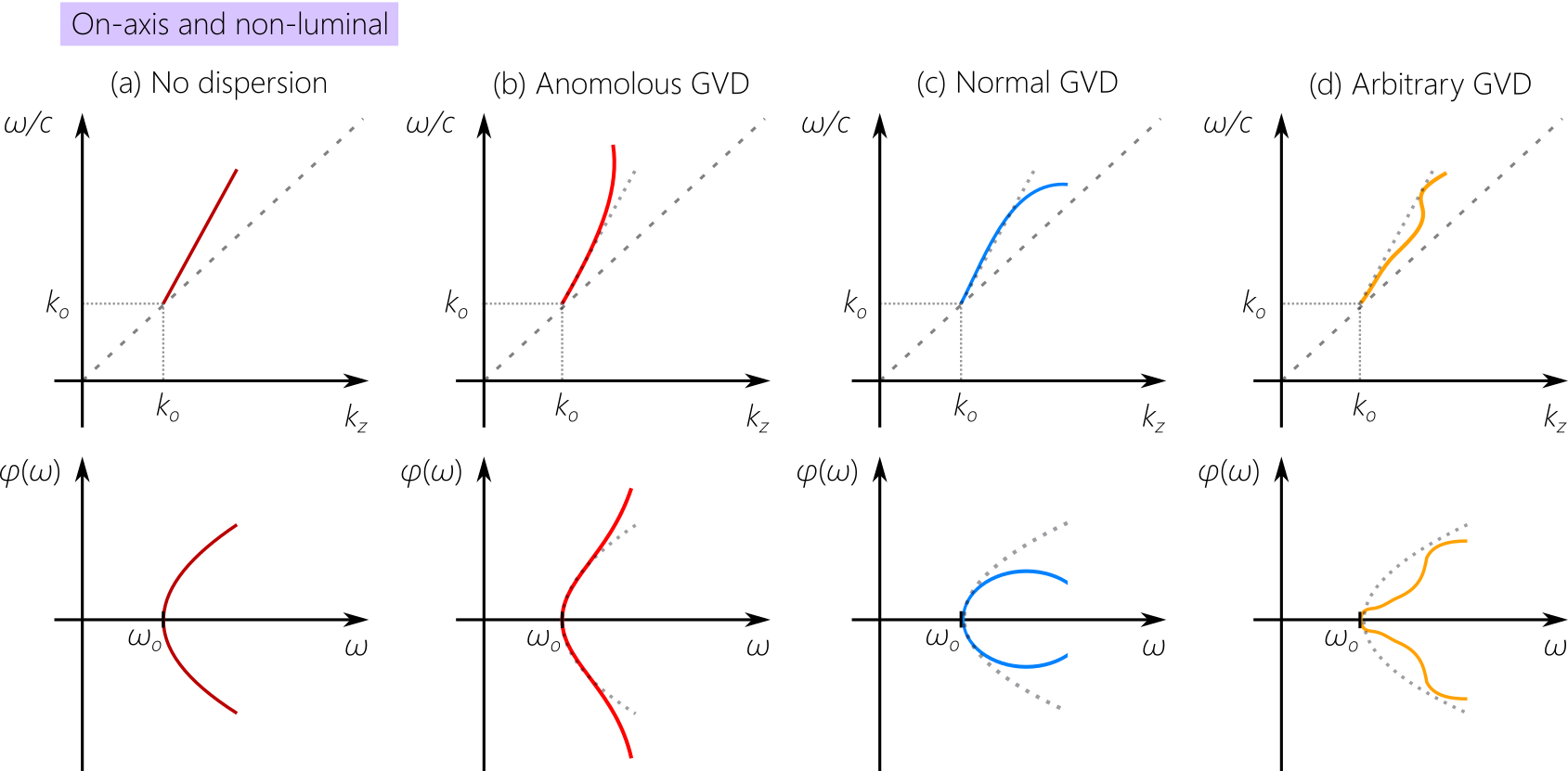}
\caption{Pulsed on-axis ($\varphi_{\mathrm{o}}\!=\!0$ and $v_{\mathrm{ph}}\!=\!c$) fields that are non-luminal ($\widetilde{v}\!\neq\!c$). (a) No dispersion; (b) anomalous GVD; (c) normal GVD; and (d) arbitrary dispersion profile. The dotted curve for $\varphi(\omega)$ in (b-d) is that for the dispersion-free case from (a). The dashed lines in the first row are the light-line $k_{z}\!=\!\tfrac{\omega}{c}$; the dotted lines in the first row are the tangents to the spectral trajectory at $\omega\!=\!\omega_{\mathrm{o}}$; and the dotted curves in the second row are $\varphi(\omega)$ for the dispersion-free field from (a).}
\label{Fig:OnAxisNonluminal}
\end{figure}

\subsection{On-axis, luminal fields}

This collection of 4 classes comprise pulsed fields propagating along the $z$-axis ($\varphi_{\mathrm{o}}\!=\!0$ and $v_{\mathrm{ph}}\!=\!c$) with luminal group velocity $\widetilde{v}\!=\!c$. The spectral trajectory in the $(k_{z},\tfrac{\omega}{c})$-plane is tangential to the light-line at $\omega\!=\!\omega_{\mathrm{o}}$, and its curvature is determined by the GVD. In absence of dispersion, the spectral trajectory lies along the light-line, and $\varphi(\omega)\!=\!0$, which is the trivial case of a plane-wave pulse traveling along the $z$-axis [Fig.~\ref{Fig:OnAxisLuminal}(a)]. In the case of anomalous GVD [Fig.~\ref{Fig:OnAxisLuminal}(b)], the spectral trajectory curves upwards away from the light-line. This wave packet has been realized in \cite{Szatmari96OL}, and can be viewed as an on-axis TPF. An on-axis luminal pulsed field endowed with \textit{nornmal} GVD is physically inadmissible because its spectral trajectory would be tangential to the light-line at $\omega\!=\!\omega_{\mathrm{o}}$ but curve away downwards. Such a field is purely evanescent [Fig.~\ref{Fig:OnAxisLuminal}(c)], and is ruled out by the result in \cite{Martinez84JOSAA}. Finally, an arbitrary dispersion profile can be inculcated as long as the spectral trajectory $k_{z}(\omega)\!=\!\tfrac{\omega}{c}+\sum_{n}\tfrac{1}{n!}k_{n}\Omega^{n}$ lies above the light-line, $k_{z}(\omega)\!<\!\tfrac{\omega}{c}$. This restriction amounts to $\sum_{n}\tfrac{1}{n!}k_{n}\Omega^{n}\!<\!0$ for all $\omega$. This can be satisfied for any magnitudes of $k_{z}^{(n)}$ as long as they are negative-valued for even-order terms, and positive-valued for odd-order terms. Having oppositely signed coefficients must be balanced against other terms in order to retain $\sum_{n}\tfrac{1}{n!}k_{n}\Omega^{n}\!<\!0$ everywhere. 

\subsection{On-axis, non-luminal fields}

This class comprises on-axis fields ($\varphi_{\mathrm{o}}\!=\!0$ and $v_{\mathrm{ph}}\!=\!c$) that are non-luminal $\widetilde{v}\!\neq\!c$, which necessitate introducing non-differentiable AD. Until recently, this class of fields went unexplored (see the theoretical studies in \cite{Valtna07OC,ZamboniRached2009PRA} for exceptions). We have recently investigated this class of pulsed fields extensively under the moniker of ST wave packets. The spectral trajectory in the $(k_{z},\tfrac{\omega}{c})$-plane must reach the light-line at $\omega\!=\!\omega_{\mathrm{o}}$ ($\varphi_{\mathrm{o}}\!=\!0$), and the tangent to the trajectory at $\omega_{\mathrm{o}}$ is \textit{not} parallel to the light-line ($\widetilde{v}\!\neq\!c$). If $\widetilde{v}\!=\!c\tan{\theta}$, then $\theta$ is the angle made by this tangent with the $k_{z}$-axis [Fig.~\ref{Fig:OnAxisNonluminal}].

The dispersion-free class [Fig.~\ref{Fig:OnAxisNonluminal}(a)] are propagation-invariant wave packets that ravel rigidly in free space at a group velocity $\widetilde{v}$. Absence of GVD implies that the spectral trajectory is a straight line making an angle $\theta$ with the $k_{z}$-axis, and changing the spectral tilt angle $\theta$ tunes $\widetilde{v}$ across the subluminal ($\theta\!<\!45^{\circ}$, $\widetilde{v}\!<\!c$), superluminal ($45^{\circ}\!<\!\theta\!<\!90^{\circ}$, $\widetilde{v}\!>\!c$), and negative-$\widetilde{v}$ ($\theta\!>\!90^{\circ}$, $\widetilde{v}\!<\!0$) regimes \cite{Kondakci19NC,Yessenov19OE,Bhaduri20NP}). Anomalous GVD [Fig.~\ref{Fig:OnAxisNonluminal}(b)] or normal GVD [Fig.~\ref{Fig:OnAxisNonluminal}(c)] can be readily introduced on equal footing making use of Eq.~\ref{Eq:AngleForArbitraryDispersion} to sculpt the necessary propagation angle $\varphi(\omega)$ that curves the spectral trajectory in the $(k_{z},\tfrac{\omega}{c})$-plane. All higher-order dispersion terms are eliminated here. Finally, realizing arbitrary GVD [Fig.~\ref{Fig:OnAxisNonluminal}(d)] is subject only to the constraint $\sum_{n=2}\tfrac{1}{n!}k_{z}^{(n)}\Omega^{n}\!<\!\big|\tfrac{\Omega}{c}(1-\widetilde{n})\big|$.

\begin{figure}[b!]
\centering
\includegraphics[width=13.2cm]{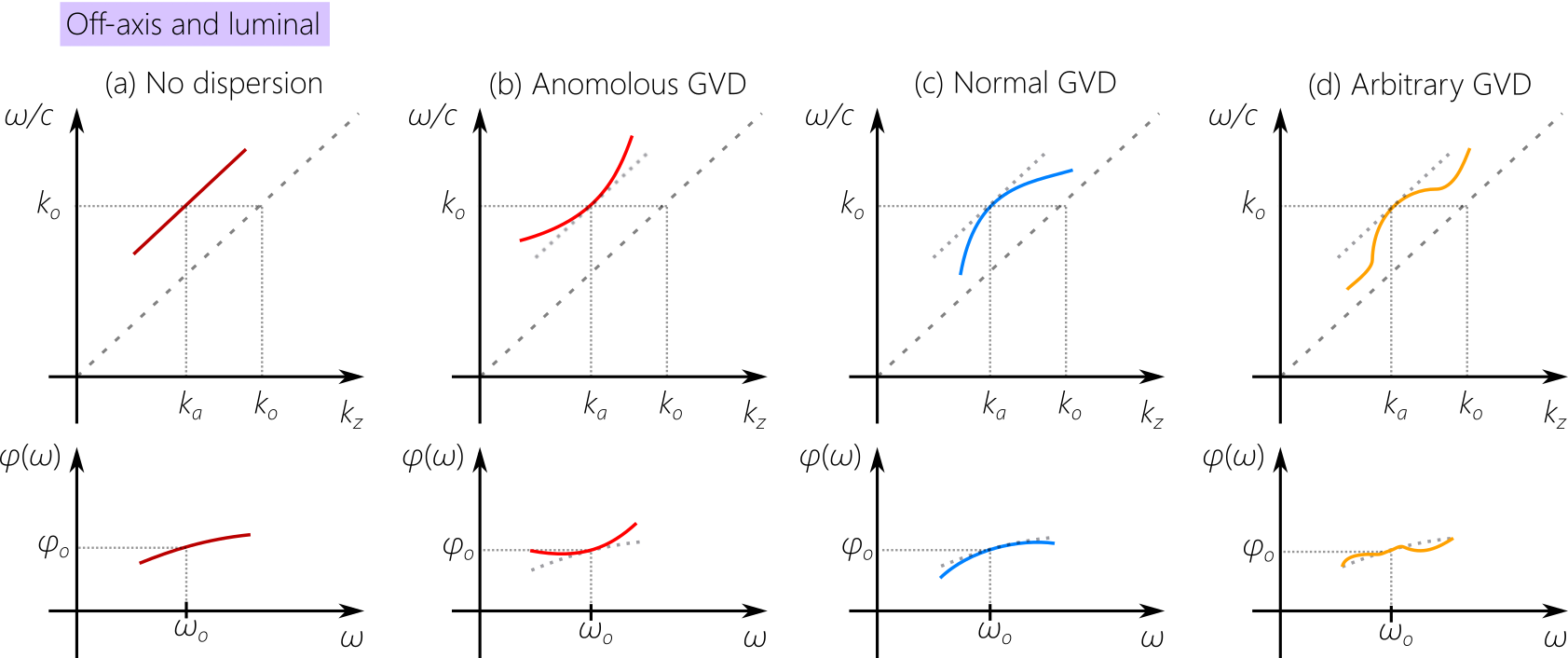}
\caption{Pulsed off-axis ($\varphi_{\mathrm{o}}\!\neq\!0$ and $v_{\mathrm{ph}}\!\neq\!c$) fields that are luminal ($\widetilde{v}\!=\!c$). (a) No dispersion; (b) anomalous GVD; (c) normal GVD; and (d) arbitrary dispersion profile. The dotted curve for $\varphi(\omega)$ in (b-d) is that for the dispersion-free case from (a). The dashed lines in the first row are the light-line $k_{z}\!=\!\tfrac{\omega}{c}$; the dotted lines in the first row are the tangents to the spectral trajectory at $\omega\!=\!\omega_{\mathrm{o}}$; and the dotted curves in the second row are $\varphi(\omega)$ for the dispersion-free field from (a).}
\label{Fig:OffAxisLuminal}
\end{figure}

\subsection{Off-axis, luminal fields}

This class comprises off-axis fields ($\varphi_{\mathrm{o}}\!\neq\!0$ and $v_{\mathrm{ph}}\!\neq\!c$) that are nevertheless luminal ($\widetilde{v}\!=\!c$) by satisfying the constraint $\varphi_{\mathrm{o}}\!=\!-2\delta^{(1)}$. Their spectral trajectories in the $(k_{z},\tfrac{\omega}{c})$-plane do \textit{not} intersect with the light-line, but the tangent to the spectral trajectory at $\omega_{\mathrm{o}}$ is parallel to the light-line. These pulsed fields do not necessitate the incorporation of non-differentiable AD.

For the dispersion-free class [Fig.~\ref{Fig:OffAxisLuminal}(a)], the spectral trajectory is a straight line parallel to the light-line but displaced with respect to it, which represents a luminal propagation-invariant wave packet. This is the focus-wave mode (FWM) discovered by Brittingham in 1983 \cite{Brittingham83JAP}, and was the first propagation-invariant wave packet identified in optics, and the only luminal one \cite{Yessenov19PRA}. By curving the spectral trajectory away from the GVD-free case, anomalous GVD can be realized [Fig.~\ref{Fig:OffAxisLuminal}(b)]. This field configuration can be produced by a grating (where GVD is \textit{always} anomalous), while satisfying $\varphi_{\mathrm{o}}\!=\!-2\delta^{(1)}$. 

In contrast to the on-axis luminal scenario where normal GVD is physically inadmissible, for the off-axis luminal fields considered here, normal GVD is indeed admissible [Fig.~\ref{Fig:OffAxisLuminal}(c)]. However, such fields have never been produced. Indeed, it can be though erroneously that the result in \cite{Martinez84JOSAA} rules such a case out. As discussed above, the theorem in \cite{Martinez84JOSAA} applies only to on-axis luminal fields. A case in point is the theoretical prediction that propagation-invariant wave packets exist in media with anomalous GVD \cite{Malaguti08OL}. The reason that normal GVD produced by AD has not been observed previously in off-axis fields is that it requires independent control over $\varphi_{\mathrm{o}}$, $\varphi_{\mathrm{o}}^{(1)}$, and $\varphi_{\mathrm{o}}^{(2)}$, which is not offered by any known optical component. 

Producing an arbitrary dispersion profile [Fig.~\ref{Fig:OffAxisLuminal}(d)] requires only that the spectral trajectory remain above the light-line. Preparing such a pulsed field requires independent control over multiple orders of AD, which has not been realized to date.

\begin{figure}[t!]
\centering
\includegraphics[width=13.2cm]{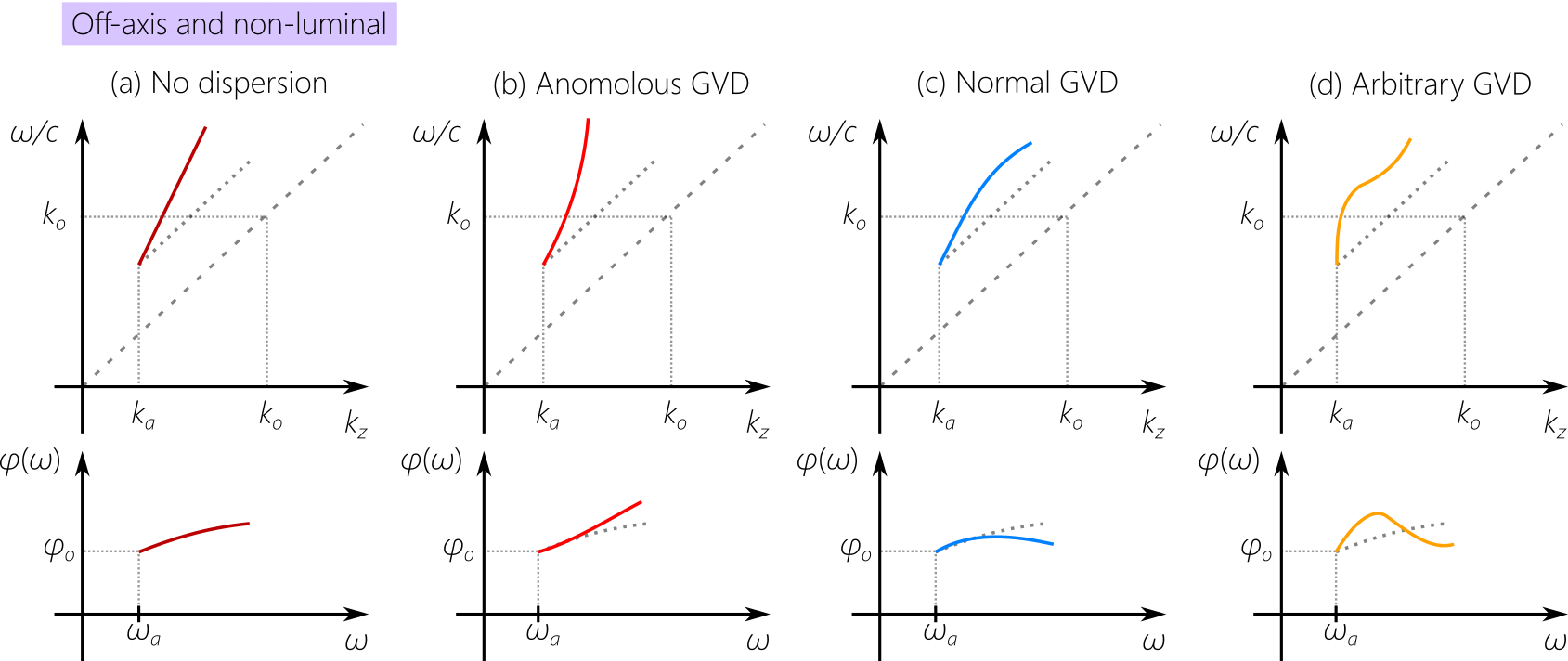}
\caption{Pulsed off-axis ($\varphi_{\mathrm{o}}\!\neq\!0$ and $v_{\mathrm{ph}}\!\neq\!c$) fields that are non-luminal ($\widetilde{v}\!\neq\!c$). (a) No dispersion; (b) anomalous GVD; (c) normal GVD; and (d) arbitrary dispersion profile. The dotted curve for $\varphi(\omega)$ in (b-d) is that for the dispersion-free case from (a). The dashed lines in the first row are the light-line $k_{z}\!=\!\tfrac{\omega}{c}$; the dotted lines in the first row are the tangents to the spectral trajectory at $\omega\!=\!\omega_{\mathrm{o}}$; and the dotted curves in the second row are $\varphi(\omega)$ for the dispersion-free field from (a).}
\label{Fig:OffAxisNonluminal}
\end{figure}

\subsection{Off-axis, non-luminal}

The spectral trajectory for an off-axis field ($\varphi_{\mathrm{o}}\!\neq\!0$ and $v_{\mathrm{ph}}\!\neq\!c$) that is non-luminal ($\widetilde{v}\!\neq\!c$) does not intersect with the light-line, and the tangent to the spectral trajectory at $\omega\!=\!\omega_{\mathrm{o}}$ is \textit{not} parallel to the light-line. Rather, this tangent makes an angle $\theta$ with the $k_{z}$-axis, such that $\widetilde{v}\!=\!c\tan{\theta}$. These 4 classes of pulsed fields do \textit{not} require non-differentiable AD for their synthesis.

In absence of dispersion [Fig.~\ref{Fig:OffAxisNonluminal}(a)], the spectral trajectory is a straight line, and thus represents a propagation-invariant wave packet with $\widetilde{v}\!\neq\!c$. This class encompasses several families of such wave packets. If the spectral trajectory when extended to low frequencies passes through the origin $k_{z}\!=\!\tfrac{\omega}{c}\!=\!0$, then it corresponds to a superluminal X-wave \cite{Lu92IEEEa,Saari97PRL}. If the extended spectral trajectory intersects with the light-line $k_{z}\!=\!-\tfrac{\omega}{c}$ ($k_{z}\!<\!0$), then it corresponds to a superluminal focused X-wave \cite{Besieris98PIERS}. Alternatively, if the extended spectral trajectory intersects with the light-line $k_{z}\!=\!\tfrac{\omega}{c}$, then it corresponds to the propagation-invariant ST wave packets in Fig.~\ref{Fig:OnAxisNonluminal}(a), except that the spectral window selected is shifted away from the non-differentiable frequency (the intersection point with the light-line). Such fields can be subluminal or superluminal, or can be even negative-values \cite{Kondakci19NC}.

Introducing anomalous GVD [Fig.~\ref{Fig:OffAxisNonluminal}(b)] can be readily done with a grating in an off-axis configuration \cite{Porras03PRE2}. This field in general is that studied as a TPF \cite{Fulop10Review,Turunen10PO}. Introducing normal GVD [Fig.~\ref{Fig:OffAxisNonluminal}(c)] requires curving the spectral trajectory toward the light-line, an example of which is the configuration proposed by Porras \textit{et al}. in \cite{Porras03PRE2}. This class requires exercising control over multiple orders of AD, which has \textit{not} been available to date. Lastly, arbitrary dispersion can be realized [Fig.~\ref{Fig:OffAxisNonluminal}(d)], once again only if control over multiple orders of AD is available.

\section{Construction for a universal angular-dispersion analyzer}

\begin{figure}[t!]
\centering
\includegraphics[width=8.6cm]{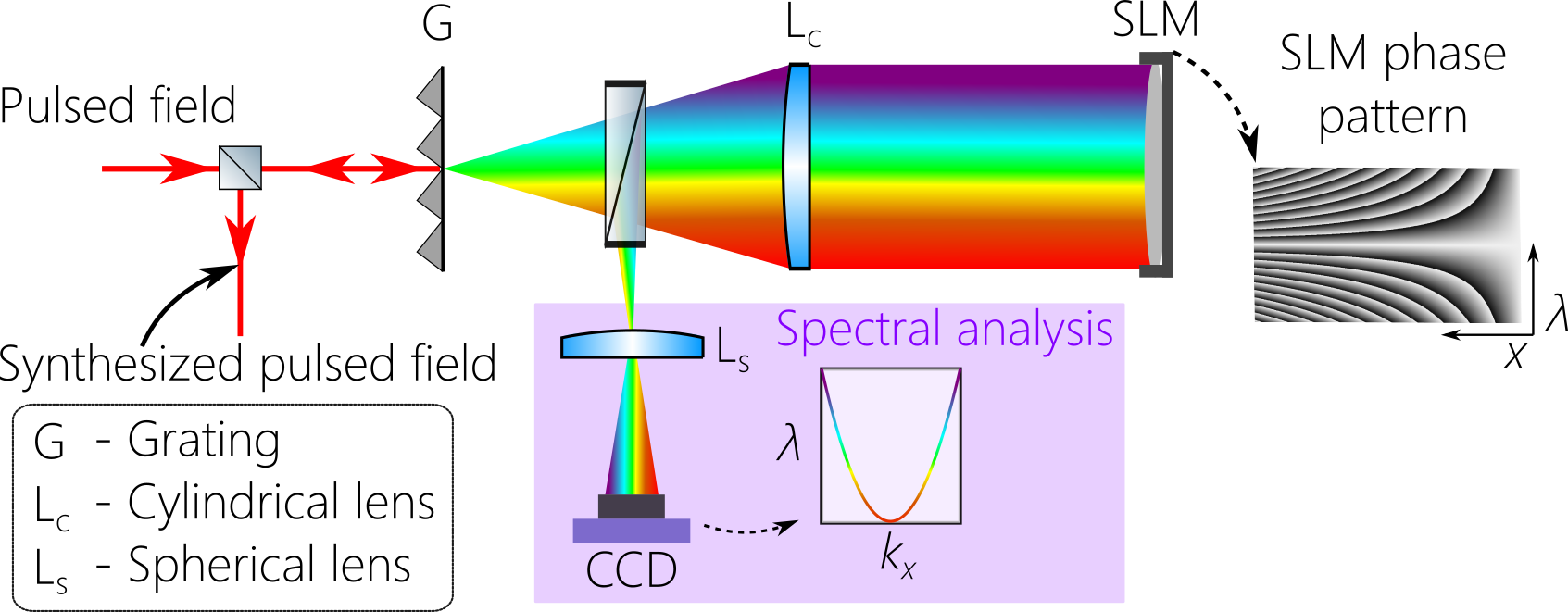}
\caption{Schematic of the optical arrangement constituting a universal angular dispersion synthesizer.}
\label{Fig:Setup}
\end{figure}

\subsection{Comparison with a conventional diffraction grating}

It is useful to examine briefly the AD produced by a conventional diffraction grating. If the incident and diffraction angles with respect to the grating normal are $\alpha$ and $\varphi$, respectively, then $\sin{\varphi}\!=\!\sin{\alpha}+m\tfrac{\lambda}{\Lambda}$, where $\lambda$ is the wavelength, $\Lambda$ is the grating period, and $m$ is the diffraction order. Only a few free parameters ($\alpha$ and $\Lambda/m$) can be tuned to modify the AD at fixed $\lambda$. Consequently, the different orders of AD are not independent of each other, which in turn entails that different dispersion orders cannot be independently tuned. For example, large values of first-order AD $\omega_{\mathrm{o}}\varphi_{\mathrm{o}}^{(1)}\!=\!(\sin{\alpha}-\sin{\varphi_{\mathrm{o}}})/\cos{\varphi_{\mathrm{o}}}$ require large $\varphi_{\mathrm{o}}$, so that $\omega_{\mathrm{o}}\varphi_{\mathrm{o}}^{(1)}$ and $\varphi_{\mathrm{o}}$ are \textit{not} independent of each other. Similarly, all higher-order AD coefficients can be written in terms of $\alpha$ and $\varphi_{\mathrm{o}}$. As such, one cannot tune one dispersion order without affecting all the others. Indeed, at $\lambda_{\mathrm{o}}\!=\!800$~nm, $\tfrac{\Delta\varphi}{\Delta\lambda}\!\sim\!3^{\circ}$/mm can only be achieved at $\varphi_{\mathrm{o}}\!\sim\!87^{\circ}$. 

\subsection{Experimental configuration}

To introduce arbitrary AD into a plane-wave pulse, we follow the two-step strategy depicted schematically in Fig.~\ref{Fig:Setup}. The first step is spectral analysis, whereby a conventional optical component (a diffraction grating here) spreads the spectrum spatially, and the spectrum is then collimated by a cylindrical lens. In the second step, the phase of the spectrally resolved spectrum in the focal plane of the lens is modulated via a SLM \cite{Kondakci17NP} or phase plate \cite{Kondakci18OE,Yessenov20OSAC}, before the spectrum is reconstituted into a pulse via a lens and a grating. In our experiment, we make use of a reflective SLM, and the retro-reflected wave front traces its steps back to the grating. Each wavelength occupies a column along the SLM. Along this direction a linear phase of the form $\Phi(\lambda)\!=\!k_{x}(\lambda)x\!=\!\tfrac{2\pi}{\lambda}\sin{\{\varphi(\lambda)\}}x$ is implemented, where $x$ is the coordinate along the column, and $\varphi(\lambda)$ is the deflection angle with respect to the $z$-axis that we aim to impart to the wavelength $\lambda$. Because the phase $\Phi$ can be set for each wavelength $\lambda$ \textit{independently}, we can produce an arbitrary angular dispersion $\varphi(\lambda)$ constrained only by the technical limits discussed below. In this way, an arbitrary functional form of $\varphi(\lambda)$ can be realized: smooth or discontinuous, differentiable or non-differentiable. For example, $\varphi(\omega)\!\propto\!\sqrt{\Omega}$ that is key to producing on-axis non-luminal pulsed fields can be readily produced. As such, this arrangement can serve as a \textit{universal AD synthesizer}.

\subsection{Large on-axis first-order AD produced by the universal AD synthesizer}

Consider the first-order dispersion coefficient $\varphi^{(1)}\!\approx\!\tfrac{\Delta\varphi}{\Delta\omega}$ where $\Delta\varphi$ is the angular spread associated with the bandwidth $\Delta\omega$. In a grating, $\Delta\varphi$ and $\Delta\lambda$ are not independent of each other, and are instead related through the grating equation. In our system, however, $\Delta\lambda$ and $\Delta\varphi$ are \textit{independent} of each other, and are controlled by two distinct processes. The reason is that we rely on the grating in our arrangement only to spatially resolve the spectrum, but \textit{not} to provide AD, so that $\Delta\omega$ is determined by the grating-lens combination, whereas $\Delta\varphi$ is determined by the numerical aperture of the SLM. With current SLM technology (pixel size $\sim\!10$~$\mu$m), the spatial resolution is far less than that of a grating (grating ruling $<\!1$~$\mu$m), one is led to expect a low $\Delta\varphi$ and thus low $\varphi^{(1)}$. Nevertheless, our system is still capable of providing large $\varphi_{\mathrm{o}}^{(1)}$. The bandwidth $\Delta\omega$ -- over which the small angular spread $\Delta\varphi$ is associated -- can be made quite small by using a high-density grating and a long-focal-length lens. By reducing $\Delta\omega$ that is incident on the SLM at fixed $\Delta\varphi$, one may obtain extremely large values of $\varphi^{(1)}$. For example, $\tfrac{d\varphi}{d\lambda}\!\sim\!3^{\circ}$/nm for $\Delta\lambda\!\sim\!1$ requires an angular spread of only $\Delta\varphi\!\approx\!3^{\circ}$.

It can be readily shown from the geometry of the problem that we have the following identity: $(\omega_{\mathrm{o}}\varphi_{\mathrm{o}}^{(1)})_{\mathrm{s}}\!=\!\tfrac{(\Delta\varphi)_{\mathrm{s}}}{(\Delta\varphi)_{\mathrm{g}}}(\omega_{\mathrm{o}}\varphi_{\mathrm{o}}^{(1)})_{\mathrm{g}}$, where the subscripts `g' and `s' indicate quantities associated with a grating or our SLM-based synthesizer, respectively. If we select a configuration where $(\Delta\varphi)_{\mathrm{s}}\!\approx\!(\Delta\varphi)_{\mathrm{g}}$, then the first-order AD produced by the two approaches are equal. Crucially, however, $(\omega_{\mathrm{o}}\varphi_{\mathrm{o}}^{(1)})_{\mathrm{s}}$ is independent of $\varphi_{\mathrm{o}}$. In fact, large values of $(\omega_{\mathrm{o}}\varphi_{\mathrm{o}}^{(1)})_{\mathrm{s}}$ can be realized on-axis rather than only at large $\varphi_{\mathrm{o}}$ as in a grating. Uniquely, arbitrary $\varphi(\omega)$ profiles can be readily synthesized, limited only by the numerical aperture, diffraction efficiency, and wavelength availability of the SLM. 

\section{Measurements}

\begin{figure}[t!]
\centering
\includegraphics[width=11cm]{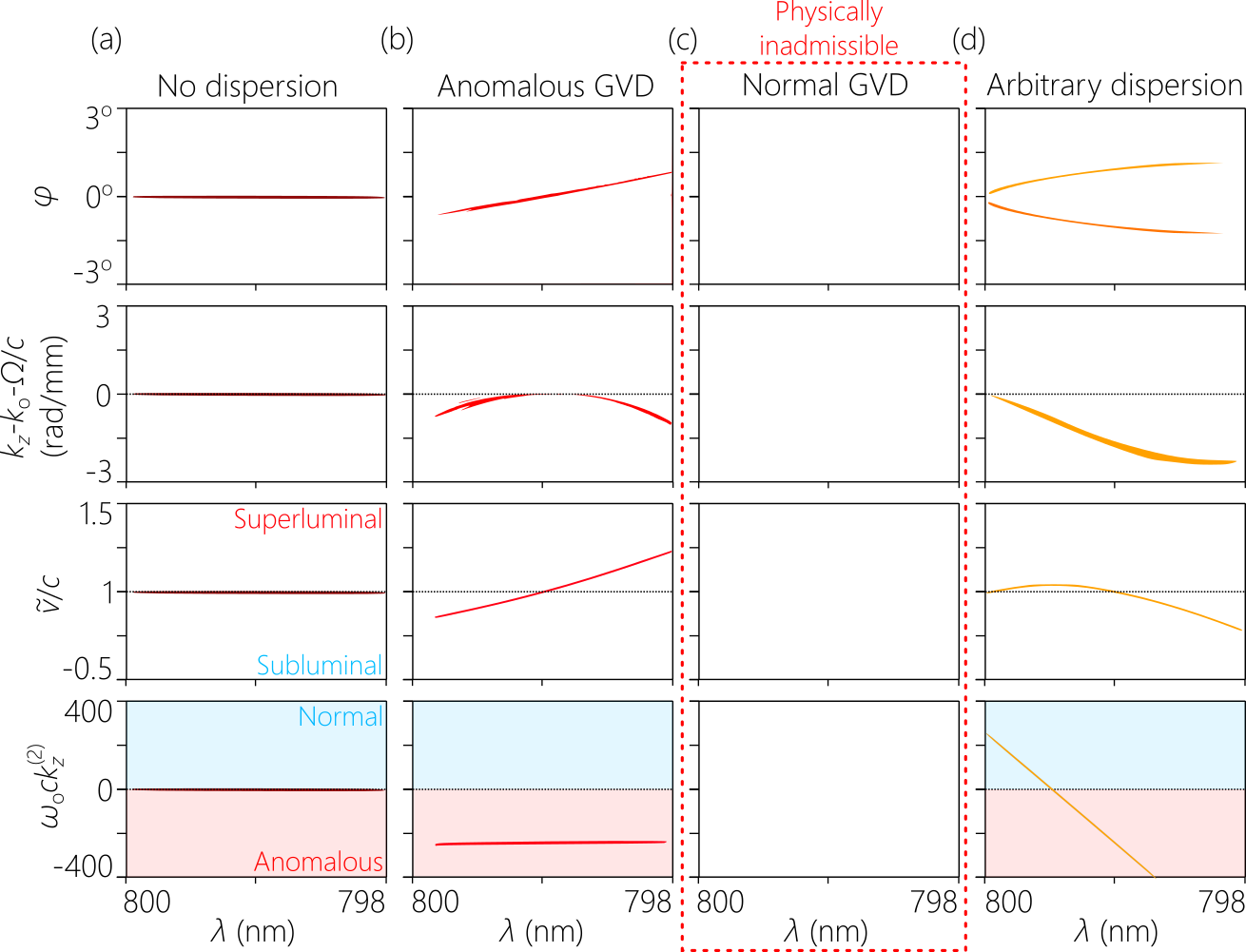}
\caption{Measurements for on-axis luminal fields. In the first row we plot $\varphi(\omega)$, in the second $k_{z}(\lambda)-k_{\mathrm{o}}-\tfrac{\Omega}{c}$, in the third $\widetilde{v}(\lambda)$, and in the fourth $k_{z}^{(2)}(\lambda)$. (a) Dispersion-free field; (b) anomalous GVD with $\omega_{\mathrm{o}}\varphi_\mathrm{o}^{(1)}\!=\!10$; (c) normal GVD is physically inadmissible; and (d) arbitrary dispersion in which $c\omega_\mathrm{o}k_{z}^{(2)}\!=\!-100$ and $c\omega_\mathrm{o}^{2}k_z^{(3)}\!=\!1.5\times10^{5}$.}
\label{Fig:Measurements1}
\end{figure}

\begin{figure}[t!]
\centering
\includegraphics[width=11cm]{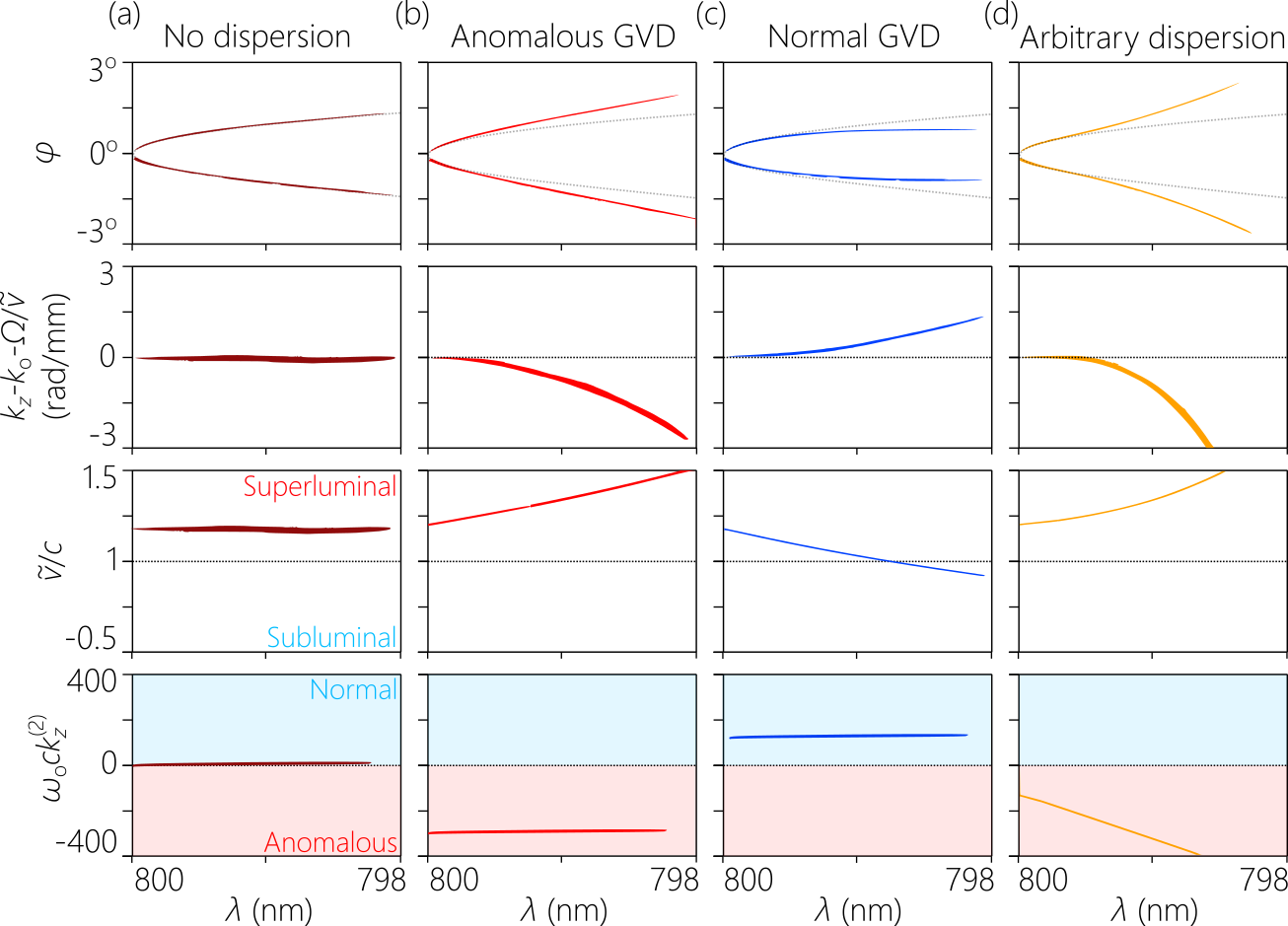}
\caption{Measurements for on-axis non-luminal fields. The rows show the same quantities as in Fig.~\ref{Fig:Measurements1} except that we plot $k_{z}-k_{\mathrm{o}}-\tfrac{\Omega}{\widetilde{v}}$ in the second. Throughout we have $\lambda_{\mathrm{o}}\!=\!800$~nm and $\widetilde{v}(\lambda_{\mathrm{o}})\!=\!1.19c$. (a) Dispersion-free field; (b) anomalous GVD with $c\omega_{\mathrm{o}}k_{z}^{(2)}\!=\!-200$; (c) normal GVD with $c\omega_{\mathrm{o}}k_{z}^{(2)}\!=\!100$; and (d) a dispersion profile in which $c\omega_\mathrm{o} k_{z}^{(2)}\!=\!-100$ and $c\omega_\mathrm{o}^{2}k_{z}^{(3)}\!=\!-10^{6}$.}
\label{Fig:Measurements2}
\end{figure}

\begin{figure}[t!]
\centering
\includegraphics[width=11cm]{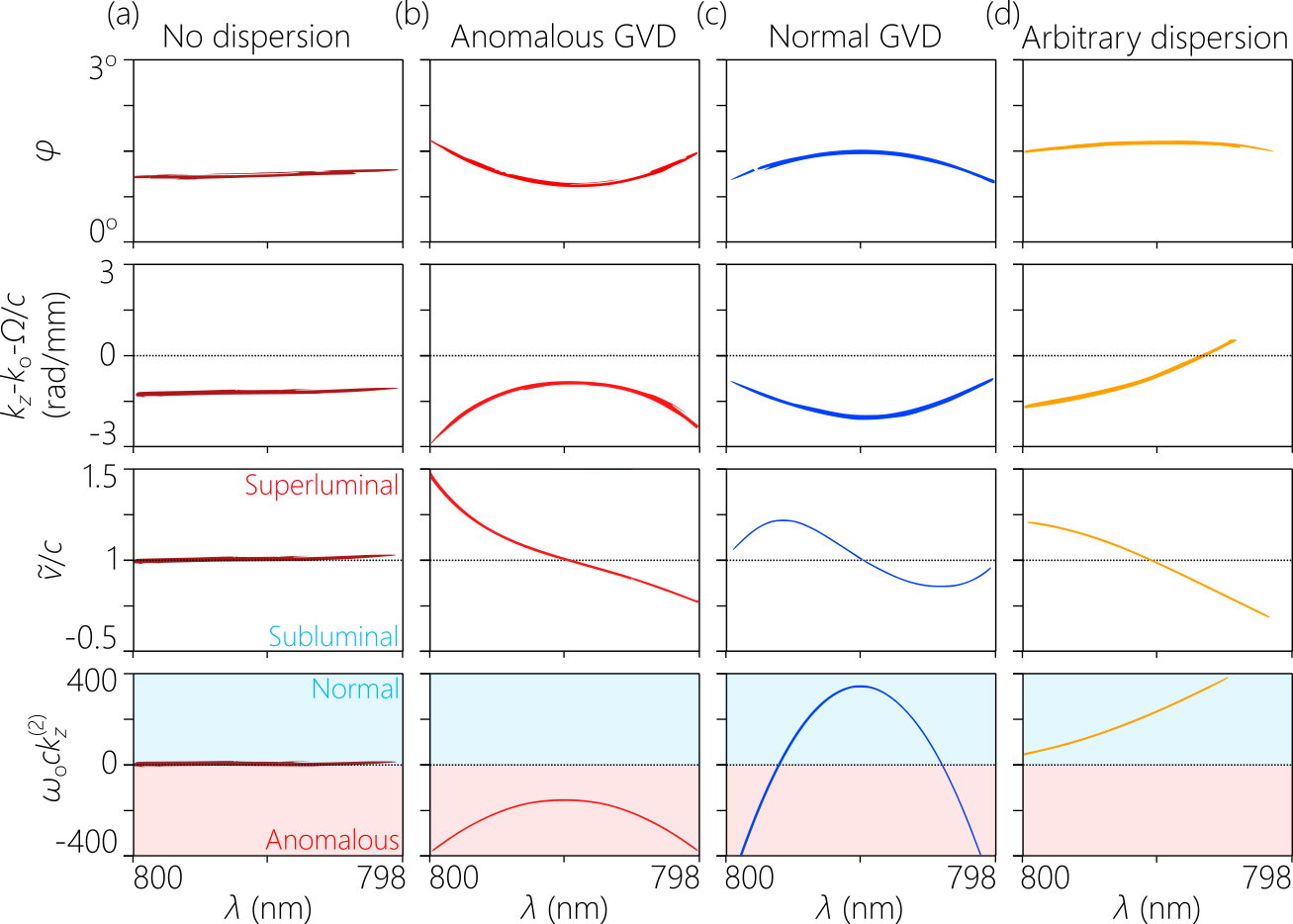}
\caption{Measurements for off-axis luminal fields with $\lambda_{\mathrm{o}}\!=\!799$~nm. The quantities plotted in the rows are the same as in Fig.~\ref{Fig:Measurements1}. (a) Dispersion-free field corresponding to a focused-wave mode with $\varphi_{\mathrm{o}}\!=\!1^{\circ}$; (b) anomalous GVD with $\varphi_{\mathrm{o}}\!=\!1^{\circ}$, $(\omega_{\mathrm{o}}\varphi_\mathrm{o}^{(1)}\!=\!0$, and $\omega_\mathrm{o}^2\varphi_\mathrm{o}^{(2)}\!=\!10^{4}$; (c) normal GVD with $\varphi_{\mathrm{o}}\!=\!1.5^{\circ}$, $(\omega_{\mathrm{o}}\varphi_\mathrm{o}^{(1)}\!=\!0$, and $\omega_\mathrm{o}^2\varphi_\mathrm{o}^{(2)}\!=\!-10^{4}$; and (d) a dispersion profile with $\varphi_{\mathrm{o}}\!=\!1.6^{\circ}$, $c\omega_\mathrm{o}k_{z}^{(2)}\!=\!50$, $c\omega_\mathrm{o}^{2}k_{z}^{(3)}\!=\!-10^{6}$, and $c\omega_\mathrm{o}^{3}k_{z}^{(4)}\!=\!5\times10^{7}$.}
\label{Fig:Measurements3}
\end{figure}

\begin{figure}[t!]
\centering
\includegraphics[width=11cm]{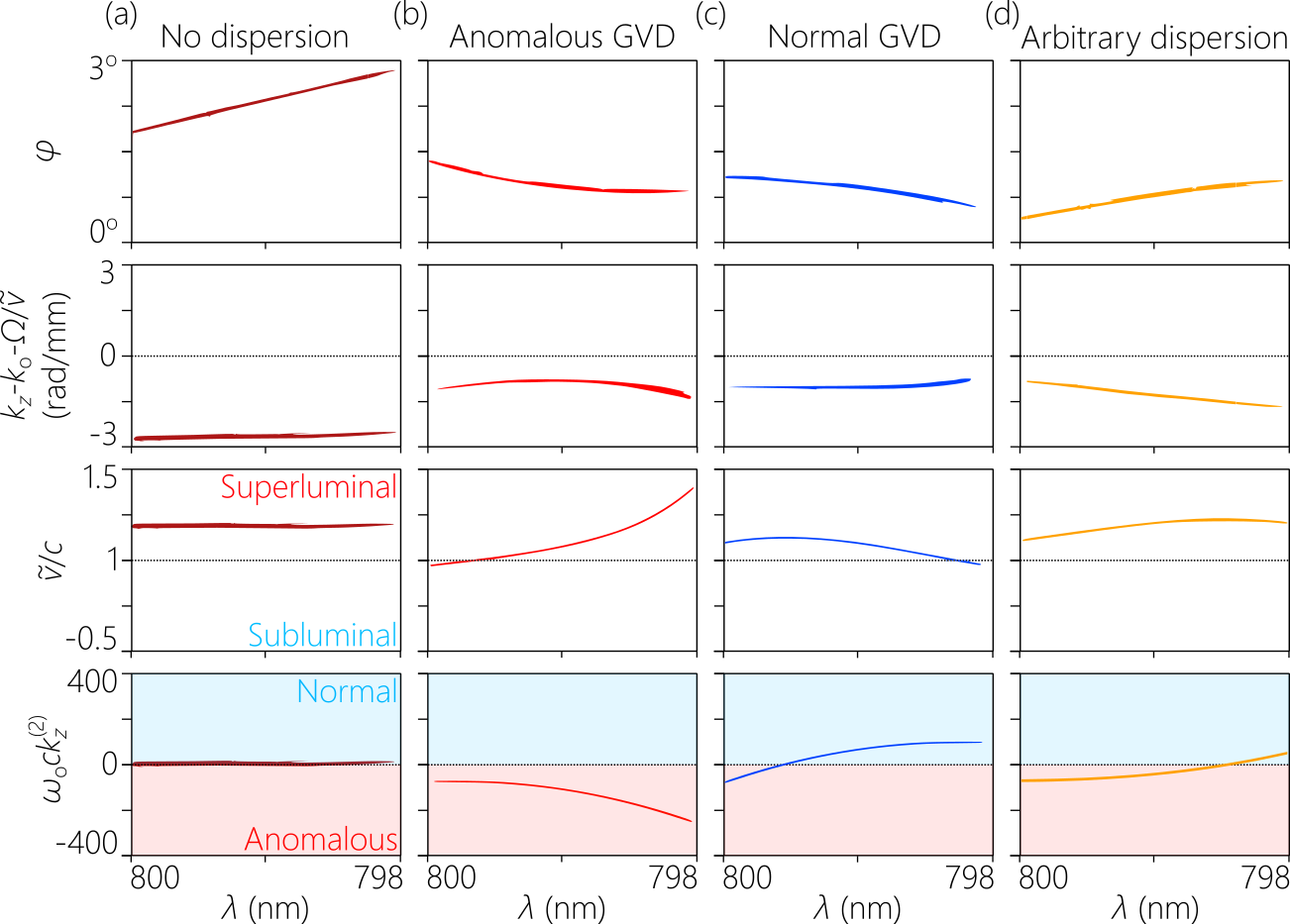}
\caption{Measurements for off-axis non-luminal fields with $\varphi_{\mathrm{o}}\!=\!1^{\circ}$ and $\lambda_{\mathrm{o}}\!=\!799$~nm. The quantities plotted in each row are the same as in Fig.~\ref{Fig:Measurements2}. (a) Dispersion-free field with $\varphi_{\mathrm{o}}\!=\!2.3^{\circ}$ and $\widetilde{v}\!=\!1.19c$; (b) anomalous GVD with $\varphi_{\mathrm{o}}\!=\!1^{\circ}$ and $\widetilde{v}(\lambda_{\mathrm{o}})\!=\!1.1c$, produced by setting $\omega_{\mathrm{o}}\varphi_\mathrm{o}^{(1)}\!=\!5$ and $\omega_\mathrm{o}^2\varphi_\mathrm{o}^{(2)}\!=\!5000$; (c) normal GVD with $\omega_{\mathrm{o}}\varphi_\mathrm{o}^{(1)}\!=\!5$ and $\omega_\mathrm{o}^2\varphi_\mathrm{o}^{(2)}\!=\!-5000$; and (d) a dispersion profile with $\varphi_{\mathrm{o}}\!=\!1^{\circ}$, $\widetilde{v}\!=\!1.19c$, $c\omega_\mathrm{o}k_{z}^{(2)}\!=\!50$, $c\omega_\mathrm{o}^{2}k_{z}^{(3)}\!=\!-10^{5}$, and $c\omega_\mathrm{o}^{3}k_{z}^{(4)}\!=\!10^{7}$.}
\label{Fig:Measurements4}
\end{figure}

Realizing any spectral profile $\varphi(\omega)$ for the propagation angle is simply a matter of implementing the requisite phase pattern on the SLM. In our experiments, we follow one of two approaches. When only a few low orders of AD are defined (say $\varphi_{\mathrm{o}}$, $\varphi_{\mathrm{o}}^{(1)}$, and $\varphi_{\mathrm{o}}^{(2)}$), and all higher orders are set to zero, then we implement on the SLM the phase $\Phi(\lambda)\!=\!k\sin{\{\varphi(\lambda)\}}$, where $\varphi(\omega)\!=\!\varphi_{\mathrm{o}}+\varphi_{\mathrm{o}}^{(1)}\Omega+\tfrac{1}{2}\varphi_{\mathrm{o}}^{(2)}\Omega^{2}$. If, on the other hand, a particular dispersion profile is targeted, say $k_{z}(\omega)\!=\!k_{\mathrm{o}}+\tfrac{\Omega}{\widetilde{v}}+\tfrac{1}{2}k_{z}^{(2)}\Omega^{2}+\tfrac{1}{6}k_{z}^{(3)}\Omega^{3}$, then we calculate $k_{x}(\omega)\!=\!\sqrt{k^{2}-k_{z}^{2}(\omega)}$, and then implement $k_{x}(\omega)$ directly.

For each field we measure $k_{x}(\omega)$ using a lens and a grating (to implement spatial and temporal Fourier transforms)  and from it we extract four quantities: (1) the frequency-dependent propagation angle $\varphi(\omega)\!=\!\arcsin{\{\tfrac{k_{x}(\omega)}{\omega/c}\}}$; (2) the axial wave number $k_{z}(\omega)\!=\!\sqrt{k^{2}-k_{x}^{2}(\omega)}\!=\!k\cos{\{\varphi(\omega)\}}$; (3) the frequency-dependent group velocity $\widetilde{v}(\omega)\!=\!\tfrac{\omega-\omega_{\mathrm{o}}}{k_{z}(\omega)-k_{\mathrm{o}}}$; and (4) the frequency-dependent GVD coefficient $k^{(2)}_{z}(\omega)\!=\!\tfrac{2}{(\omega-\omega_{\mathrm{o}})^{2}}\{k_{z}(\omega)-k_{\mathrm{o}}-\tfrac{\omega-\omega_{\mathrm{o}}}{\widetilde{v}(\omega_{\mathrm{o}})}\}$. For luminal fields (on-axis [Fig.~\ref{Fig:Measurements1}] or off-axis [Fig.~\ref{Fig:Measurements3}]), we plot $k_{z}(\omega)$ relative to the light-line, $k_{z}(\omega)-k_{\mathrm{o}}-\tfrac{\Omega}{c}$, for which only negative values are allowed. For non-luminal fields (on-axis [Fig.~\ref{Fig:Measurements2}] or off-axis [Fig.~\ref{Fig:Measurements4}]), we plot the more convenient quantity $k_{z}(\omega)-k_{\mathrm{o}}-\tfrac{\Omega}{\widetilde{v}}$, which may be either positive or negative. 

We systematically realize examples of all 15 physically admissible field configurations enumerated in Fig.~\ref{Fig:Classification}, as we proceed to show.

First, we plot the measurement results for the on-axis luminal fields in Fig.~\ref{Fig:Measurements1} with $\lambda_{\mathrm{o}}\!=\!799$~nm, whereupon $\widetilde{v}(\lambda_{\mathrm{o}})\!=\!c$ and $\varphi(\lambda_{\mathrm{o}})\!=\!\varphi_{\mathrm{o}}\!=\!0$. For the dispersion-free field, $k_{z}(\omega)\!=\!\frac{\omega}{c}$ and $\varphi(\omega)\!=\!0$, corresponding to a plane-wave pulse with $\widetilde{v}(\omega)\!=\!c$ and $k_{z}^{(2)}(\omega)\!=\!0$, as shown in Fig.~\ref{Fig:Measurements1}(a). We include anomalous GVD by adding first-order AD $\omega_{\mathrm{o}}\varphi_{\mathrm{o}}^{(1)}\!=\!10$, which leads to $\varphi(\omega)\!\propto\!\omega$, a wavelength-dependent $\widetilde{v}$, and negative $k_{z}^{(2)}$ [Fig.~\ref{Fig:Measurements1}(b)]. The normal GVD class here is physically inadmissible [Fig.~\ref{Fig:Measurements1}(b)]. In Fig.~\ref{Fig:Measurements1}(d) we show a field whose second- and third-order dispersion coefficients, $k_{z}^{(2)}$ and $k_{z}^{(3)}$, have been set at negative and positive values, respectively. The latter field was synthesized making use of non-differentiable AD to introduce the prescribed on-axis dispersion profile.

Second, for on-axis non-luminal fields [Fig.~\ref{Fig:Measurements2}], the AD must be non-differentiable, with $\lambda_{\mathrm{o}}\!=\!800$~nm, a superluminal group velocity $\widetilde{v}(\lambda_{\mathrm{o}})\!=\!1.19c$, and $\varphi(\lambda_{\mathrm{o}})\!=\!\varphi_{\mathrm{o}}\!=\!0$ throughout. The dispersion-free case is a propagation-invariant ST wave packet \cite{Kondakci17NP,Yessenov19OPN} with $\varphi(\omega)\!\propto\!\sqrt{\Omega}$, $\widetilde{v}(\omega)\!=\!1.19c$ independently of $\omega$, and vanishing dispersion of all orders, as shown in Fig.~\ref{Fig:Measurements2}(a). Anomalous GVD [Fig.~\ref{Fig:Measurements2}(b)] or normal GVD [Fig.~\ref{Fig:Measurements2}(c)] are produced using $\varphi(\omega)$ from Eq.~\ref{Eq:AngleForArbitraryDispersion} after setting $k_{2}$ to negative or positive values, respectively. In both cases, $\widetilde{v}(\omega)$ becomes frequency-dependent with $\widetilde{v}(\lambda_{\mathrm{o}})\!=\!1.19c$, whereas $k_{z}^{(2)}$ is frequency-independent, thus signifying that all dispersion orders above second are eliminated. Finally, a general dispersion profile is shown in Fig.~\ref{Fig:Measurements2}(d) where we set the values of $k_{z}^{(2)}$ and $k_{z}^{(3)}$.

The classes belonging to the third category of off-axis luminal fields are presented in Fig.~\ref{Fig:Measurements3}. In all cases, we guarantee that $\widetilde{v}(\lambda_{\mathrm{o}})\!=\!c$ by setting $\varphi_{\mathrm{o}}\!=\!-2\delta_{\mathrm{o}}^{(1)}$, or alternatively $\tan{\varphi_{\mathrm{o}}}\!=\!-2\tfrac{\omega_{\mathrm{o}}\varphi_{\mathrm{o}}^{(1)}}{1-(\omega_{\mathrm{o}}\varphi_{\mathrm{o}}^{(1)})^{2}}$; i.e., we implement a particular relationship between $\varphi_{\mathrm{o}}$ and $\varphi_{\mathrm{o}}^{(1)}$. In the dispersion-free case [Fig.~\ref{Fig:Measurements3}(a)], $\varphi(\omega)\!=\!\varphi_{\mathrm{o}}$, $\widetilde{v}(\omega)\!=\!c$, and $k_{z}^{(2)}\!=\!0$. This propagation-invariant pulsed field is a segment from a focused-wave mode. Because of the limited bandwidth, however, the characteristic X-shaped spatio-temporal profile would not be visible \cite{Reivelt02PRE}. We introduce anomalous GVD [Fig.~\ref{Fig:Measurements3}(b)] and normal GVD [Fig.~\ref{Fig:Measurements3}(c)] by setting $\varphi_{\mathrm{o}}^{(1)}\!=\!0$ and changing the sign of a non-zero $\varphi_{\mathrm{o}}^{(2)}$. In Fig.~\ref{Fig:Measurements3}(d) we plot the results for a field synthesized with a dispersion profile in which the values of $k_{z}^{(2)}$, $k_{z}^{(3)}$, and $k_{z}^{(4)}$ are set.

Finally, we plot the measurements for off-axis non-luminal fields in Fig.~\ref{Fig:Measurements4}, where the AD introduced is also differentiable. By removing the constraint $\varphi_{\mathrm{o}}\!=\!-2\delta_{\mathrm{o}}^{(1)}$, we now have $\widetilde{v}(\lambda_{\mathrm{o}})\!\neq\!c$, where $\lambda_{\mathrm{o}}\!=\!799$~nm. For the dispersion-free field [Fig.~\ref{Fig:Measurements4}(a)], $\varphi(\omega)$ is non-differentiable and corresponds to a ST wave packet with $\widetilde{v}\!=\!1.19c$, but we are away from its on-axis non-differentiable wavelength, and the entire spectrum considered here is off-axis. Consequently, $\widetilde{v}(\omega)\!=\!1.19c$ and $k_{z}^{(2)}\!=\!0$. We include anomalous GVD [Fig.~\ref{Fig:Measurements4}(b)] or normal GVD [Fig.~\ref{Fig:Measurements4}(c)] by adjusting the values of $\varphi_{\mathrm{o}}$, $\varphi_{\mathrm{o}}^{(1)}$, and $\varphi_{\mathrm{o}}^{(2)}$ in Eq.~\ref{Eq:SecondOrderDispersionTerm} independently. Note that $k_{z}^{(2)}(\omega)$ in these two scenarios are wavelength-dependent, indicating that higher-order dispersion terms are \textit{not} negligible. Lastly, we present in Fig.~\ref{Fig:Measurements4}(d) a field whose dispersion profile has the values of $k_{z}^{(2)}$, $k_{z}^{(3)}$, and $k_{z}^{(4)}$ are all set.

\section{Discussion and conclusion}

One virtue of this systematic survey is that some field configurations which might perhaps otherwise have escaped attention can be instead identified and examined. There is of course an element of arbitrariness in this classification. One could further subdivide the non-luminal fields into sub-categories: subluminal, superluminal, and negative-$\widetilde{v}$ regimes. Although convincing arguments could be mounted supporting this more detailed classification, we nevertheless group all the non-luminal fields together because the major challenge from the experimental perspective lies in the distinction between the luminal and non-luminal cases. Such a subdivision would increase the fraction of fields that have \textit{not} yet been produced, so that our more restricted classification scheme estimates the fraction of yet-to-be realized fields more conservatively.

In our work, we have made use of bulk optical components to construct the universal AD synthesizer. It is important to explore other potential platforms for realizing the same capability. Prime candidates include free-form optics, volume grating systems, metasurfaces, and nanophotonic devices. It is a sobering thought that despite tremendous progress in various aspects of nanophotonics over the past few decades, no known device can produce the missing classes of pulsed fields in Fig.~\ref{Fig:Classification}. One exception is a recent theoretical proposal for the synthesis of on-axis non-luminal pulsed optical fields (propagation-invariant ST wave packets) via a non-local nanophotonic device \cite{Guo21Light}. Our work therefore points to a potential role for nanophotonics in modulating pulsed optical fields by producing controllable AD.

The experimental arrangement described here is capable of introducing AD in one transverse spatial coordinate, but not the other. The main constraint arises from the use of a 2D SLM in which one dimension is reserved for modulating the temporal spectrum, therefore leaving only one dimension for modulating the field spatially. Conventional techniques for producing AD using prisms, gratings, or other dispersive or diffractive devices all introduce AD in one dimension. It remains an open question whether it is possible to construct a universal AD synthesizer in two transverse dimensions.

In conclusion, we have described an optical arrangement that serves as a versatile, high-resolution, \textit{universal angular-dispersion synthesizer} capable of inculcating an arbitrary wavelength-dependent propagation angle into a pulsed optical field. Using this system, we realize representative examples from all 15 physically admissible classes of fields from the 16 possible classes categorized by axial phase velocity, group velocity, and dispersion profile. Access to this broad span of structured pulsed fields with precisely tailored dispersion profiles can help improve nonlinear interactions with materials or structures in the vicinity of their resonances where the refractive index changes rapidly. Furthermore, such sculpted fields will benefit investigations of novel guided pulsed modes \cite{Shiri20NC,Guo21PRR}, and omni-resonant interactions with planar cavities \cite{Shiri20OL,Shiri20APLP}.

\bibliography{diffraction}

\begin{thebibliography}{69}%
\makeatletter
\providecommand \@ifxundefined [1]{%
 \@ifx{#1\undefined}
}%
\providecommand \@ifnum [1]{%
 \ifnum #1\expandafter \@firstoftwo
 \else \expandafter \@secondoftwo
 \fi
}%
\providecommand \@ifx [1]{%
 \ifx #1\expandafter \@firstoftwo
 \else \expandafter \@secondoftwo
 \fi
}%
\providecommand \natexlab [1]{#1}%
\providecommand \enquote  [1]{``#1''}%
\providecommand \bibnamefont  [1]{#1}%
\providecommand \bibfnamefont [1]{#1}%
\providecommand \citenamefont [1]{#1}%
\providecommand \href@noop [0]{\@secondoftwo}%
\providecommand \href [0]{\begingroup \@sanitize@url \@href}%
\providecommand \@href[1]{\@@startlink{#1}\@@href}%
\providecommand \@@href[1]{\endgroup#1\@@endlink}%
\providecommand \@sanitize@url [0]{\catcode `\\12\catcode `\$12\catcode
  `\&12\catcode `\#12\catcode `\^12\catcode `\_12\catcode `\%12\relax}%
\providecommand \@@startlink[1]{}%
\providecommand \@@endlink[0]{}%
\providecommand \url  [0]{\begingroup\@sanitize@url \@url }%
\providecommand \@url [1]{\endgroup\@href {#1}{\urlprefix }}%
\providecommand \urlprefix  [0]{URL }%
\providecommand \Eprint [0]{\href }%
\providecommand \doibase [0]{https://doi.org/}%
\providecommand \selectlanguage [0]{\@gobble}%
\providecommand \bibinfo  [0]{\@secondoftwo}%
\providecommand \bibfield  [0]{\@secondoftwo}%
\providecommand \translation [1]{[#1]}%
\providecommand \BibitemOpen [0]{}%
\providecommand \bibitemStop [0]{}%
\providecommand \bibitemNoStop [0]{.\EOS\space}%
\providecommand \EOS [0]{\spacefactor3000\relax}%
\providecommand \BibitemShut  [1]{\csname bibitem#1\endcsname}%
\let\auto@bib@innerbib\@empty
\bibitem [{\citenamefont {Sabra}(1981)}]{Sabra81Book}%
  \BibitemOpen
  \bibfield  {author} {\bibinfo {author} {\bibfnamefont {A.~I.}\ \bibnamefont
  {Sabra}},\ }\href@noop {} {\emph {\bibinfo {title} {Theories of Light from
  Descartes to Newton}}}\ (\bibinfo  {publisher} {Cambridge Univ. Press},\
  \bibinfo {year} {1981})\BibitemShut {NoStop}%
\bibitem [{\citenamefont {F{\"u}l{\"o}p}\ and\ \citenamefont
  {Hebling}(2010)}]{Fulop10Review}%
  \BibitemOpen
  \bibfield  {author} {\bibinfo {author} {\bibfnamefont {J.~A.}\ \bibnamefont
  {F{\"u}l{\"o}p}}\ and\ \bibinfo {author} {\bibfnamefont {J.}~\bibnamefont
  {Hebling}},\ }\bibfield  {title} {\bibinfo {title} {Applications of
  tilted-pulse-front excitation},\ }in\ \href@noop {} {\emph {\bibinfo
  {booktitle} {Recent Optical and Photonic Technologies}}},\ \bibinfo {editor}
  {edited by\ \bibinfo {editor} {\bibfnamefont {K.~Y.}\ \bibnamefont {Kim}}}\
  (\bibinfo  {publisher} {InTech},\ \bibinfo {year} {2010})\BibitemShut
  {NoStop}%
\bibitem [{\citenamefont {Torres}\ \emph {et~al.}(2010)\citenamefont {Torres},
  \citenamefont {Hendrych},\ and\ \citenamefont {Valencia}}]{Torres10AOP}%
  \BibitemOpen
  \bibfield  {author} {\bibinfo {author} {\bibfnamefont {J.~P.}\ \bibnamefont
  {Torres}}, \bibinfo {author} {\bibfnamefont {M.}~\bibnamefont {Hendrych}},\
  and\ \bibinfo {author} {\bibfnamefont {A.}~\bibnamefont {Valencia}},\
  }\bibfield  {title} {\bibinfo {title} {Angular dispersion: an enabling tool
  in nonlinear and quantum optics},\ }\href@noop {} {\bibfield  {journal}
  {\bibinfo  {journal} {Adv. Opt. Photon.}\ }\textbf {\bibinfo {volume} {2}},\
  \bibinfo {pages} {319} (\bibinfo {year} {2010})}\BibitemShut {NoStop}%
\bibitem [{\citenamefont {Hebling}\ \emph {et~al.}(2002)\citenamefont
  {Hebling}, \citenamefont {Almási}, \citenamefont {Kozma},\ and\
  \citenamefont {Kuhl}}]{Hebling02OE}%
  \BibitemOpen
  \bibfield  {author} {\bibinfo {author} {\bibfnamefont {J.}~\bibnamefont
  {Hebling}}, \bibinfo {author} {\bibfnamefont {G.}~\bibnamefont {Almási}},
  \bibinfo {author} {\bibfnamefont {I.~Z.}\ \bibnamefont {Kozma}},\ and\
  \bibinfo {author} {\bibfnamefont {J.}~\bibnamefont {Kuhl}},\ }\bibfield
  {title} {\bibinfo {title} {Velocity matching by pulse front tilting for
  large-area {TH}z-pulse generation},\ }\href@noop {} {\bibfield  {journal}
  {\bibinfo  {journal} {Opt. Express}\ }\textbf {\bibinfo {volume} {10}},\
  \bibinfo {pages} {1161} (\bibinfo {year} {2002})}\BibitemShut {NoStop}%
\bibitem [{\citenamefont {Szatm{\'a}ri}\ \emph {et~al.}(1996)\citenamefont
  {Szatm{\'a}ri}, \citenamefont {Simon},\ and\ \citenamefont
  {Feuerhake}}]{Szatmari96OL}%
  \BibitemOpen
  \bibfield  {author} {\bibinfo {author} {\bibfnamefont {S.}~\bibnamefont
  {Szatm{\'a}ri}}, \bibinfo {author} {\bibfnamefont {P.}~\bibnamefont
  {Simon}},\ and\ \bibinfo {author} {\bibfnamefont {M.}~\bibnamefont
  {Feuerhake}},\ }\bibfield  {title} {\bibinfo {title}
  {Group-velocity-dispersion-compensated propagation of short pulses in
  dispersive media},\ }\href@noop {} {\bibfield  {journal} {\bibinfo  {journal}
  {Opt. Lett.}\ }\textbf {\bibinfo {volume} {21}},\ \bibinfo {pages} {1156}
  (\bibinfo {year} {1996})}\BibitemShut {NoStop}%
\bibitem [{\citenamefont {Martinez}\ \emph {et~al.}(1984)\citenamefont
  {Martinez}, \citenamefont {Gordon},\ and\ \citenamefont
  {Fork}}]{Martinez84JOSAA}%
  \BibitemOpen
  \bibfield  {author} {\bibinfo {author} {\bibfnamefont {O.~E.}\ \bibnamefont
  {Martinez}}, \bibinfo {author} {\bibfnamefont {J.~P.}\ \bibnamefont
  {Gordon}},\ and\ \bibinfo {author} {\bibfnamefont {R.~L.}\ \bibnamefont
  {Fork}},\ }\bibfield  {title} {\bibinfo {title} {Negative group-velocity
  dispersion using refraction},\ }\href@noop {} {\bibfield  {journal} {\bibinfo
   {journal} {J. Opt. Soc. Am. A}\ }\textbf {\bibinfo {volume} {1}},\ \bibinfo
  {pages} {1003} (\bibinfo {year} {1984})}\BibitemShut {NoStop}%
\bibitem [{\citenamefont {Fork}\ \emph {et~al.}(1984)\citenamefont {Fork},
  \citenamefont {Martinez},\ and\ \citenamefont {Gordon}}]{Fork84OL}%
  \BibitemOpen
  \bibfield  {author} {\bibinfo {author} {\bibfnamefont {R.~L.}\ \bibnamefont
  {Fork}}, \bibinfo {author} {\bibfnamefont {O.~E.}\ \bibnamefont {Martinez}},\
  and\ \bibinfo {author} {\bibfnamefont {J.~P.}\ \bibnamefont {Gordon}},\
  }\bibfield  {title} {\bibinfo {title} {Negative dispersion using pairs of
  prisms},\ }\href@noop {} {\bibfield  {journal} {\bibinfo  {journal} {Opt.
  Lett.}\ }\textbf {\bibinfo {volume} {9}},\ \bibinfo {pages} {150} (\bibinfo
  {year} {1984})}\BibitemShut {NoStop}%
\bibitem [{\citenamefont {Gordon}\ and\ \citenamefont
  {Fork}(1984)}]{Gordon84OL}%
  \BibitemOpen
  \bibfield  {author} {\bibinfo {author} {\bibfnamefont {J.~P.}\ \bibnamefont
  {Gordon}}\ and\ \bibinfo {author} {\bibfnamefont {R.~L.}\ \bibnamefont
  {Fork}},\ }\bibfield  {title} {\bibinfo {title} {Optical resonator with
  negative dispersion},\ }\href@noop {} {\bibfield  {journal} {\bibinfo
  {journal} {Opt. Lett.}\ }\textbf {\bibinfo {volume} {9}},\ \bibinfo {pages}
  {153} (\bibinfo {year} {1984})}\BibitemShut {NoStop}%
\bibitem [{\citenamefont {Bor}\ and\ \citenamefont {R{\'a}cz}(1985)}]{Bor85OC}%
  \BibitemOpen
  \bibfield  {author} {\bibinfo {author} {\bibfnamefont {Z.}~\bibnamefont
  {Bor}}\ and\ \bibinfo {author} {\bibfnamefont {B.}~\bibnamefont {R{\'a}cz}},\
  }\bibfield  {title} {\bibinfo {title} {Group velocity dispersion in prisms
  and its application to pulse compression and travelling-wave excitation},\
  }\href@noop {} {\bibfield  {journal} {\bibinfo  {journal} {Opt. Commun.}\
  }\textbf {\bibinfo {volume} {54}},\ \bibinfo {pages} {165} (\bibinfo {year}
  {1985})}\BibitemShut {NoStop}%
\bibitem [{\citenamefont {Lemoff}\ and\ \citenamefont
  {Barty}(1993)}]{Lemoff93OL}%
  \BibitemOpen
  \bibfield  {author} {\bibinfo {author} {\bibfnamefont {B.~E.}\ \bibnamefont
  {Lemoff}}\ and\ \bibinfo {author} {\bibfnamefont {C.~P.~J.}\ \bibnamefont
  {Barty}},\ }\bibfield  {title} {\bibinfo {title} {Quintic-phase-limited,
  spatially uniform expansion and recompression of ultrashort optical pulses},\
  }\href@noop {} {\bibfield  {journal} {\bibinfo  {journal} {Opt. Lett.}\
  }\textbf {\bibinfo {volume} {18}},\ \bibinfo {pages} {1651} (\bibinfo {year}
  {1993})}\BibitemShut {NoStop}%
\bibitem [{\citenamefont {Kane}\ and\ \citenamefont
  {Squier}(1997)}]{Kane97JOSAB}%
  \BibitemOpen
  \bibfield  {author} {\bibinfo {author} {\bibfnamefont {S.}~\bibnamefont
  {Kane}}\ and\ \bibinfo {author} {\bibfnamefont {J.}~\bibnamefont {Squier}},\
  }\bibfield  {title} {\bibinfo {title} {Grism-pair stretcher--compressor
  system for simultaneous second- and third-order dispersion compensation in
  chirped-pulse amplification},\ }\href@noop {} {\bibfield  {journal} {\bibinfo
   {journal} {J. Opt. Soc. Am. B}\ }\textbf {\bibinfo {volume} {14}},\ \bibinfo
  {pages} {661} (\bibinfo {year} {1997})}\BibitemShut {NoStop}%
\bibitem [{\citenamefont {Martinez}(1989)}]{Martinez89IEEE}%
  \BibitemOpen
  \bibfield  {author} {\bibinfo {author} {\bibfnamefont {O.~E.}\ \bibnamefont
  {Martinez}},\ }\bibfield  {title} {\bibinfo {title} {Achromatic phase
  matching for second harmonic generation of femtosecond pulses},\ }\href@noop
  {} {\bibfield  {journal} {\bibinfo  {journal} {IEEE J. Sel. Top. Quantum
  Electron.}\ }\textbf {\bibinfo {volume} {25}},\ \bibinfo {pages} {2464}
  (\bibinfo {year} {1989})}\BibitemShut {NoStop}%
\bibitem [{\citenamefont {Szab{\'o}}\ and\ \citenamefont
  {Bor}(1990)}]{Szabo90APB}%
  \BibitemOpen
  \bibfield  {author} {\bibinfo {author} {\bibfnamefont {G.}~\bibnamefont
  {Szab{\'o}}}\ and\ \bibinfo {author} {\bibfnamefont {Z.}~\bibnamefont
  {Bor}},\ }\bibfield  {title} {\bibinfo {title} {Broadband frequency doubler
  for femtosecond pulses},\ }\href@noop {} {\bibfield  {journal} {\bibinfo
  {journal} {Appl. Phys. B}\ }\textbf {\bibinfo {volume} {50}},\ \bibinfo
  {pages} {51} (\bibinfo {year} {1990})}\BibitemShut {NoStop}%
\bibitem [{\citenamefont {Szab{\'o}}\ and\ \citenamefont
  {Bor}(1994)}]{Szabo94APB}%
  \BibitemOpen
  \bibfield  {author} {\bibinfo {author} {\bibfnamefont {G.}~\bibnamefont
  {Szab{\'o}}}\ and\ \bibinfo {author} {\bibfnamefont {Z.}~\bibnamefont
  {Bor}},\ }\bibfield  {title} {\bibinfo {title} {Frequency conversion of
  ultrashort pulses},\ }\href@noop {} {\bibfield  {journal} {\bibinfo
  {journal} {Appl. Phys. B}\ }\textbf {\bibinfo {volume} {58}},\ \bibinfo
  {pages} {237} (\bibinfo {year} {1994})}\BibitemShut {NoStop}%
\bibitem [{\citenamefont {Richman}\ \emph {et~al.}(1998)\citenamefont
  {Richman}, \citenamefont {Bisson}, \citenamefont {Trebino}, \citenamefont
  {Sidick},\ and\ \citenamefont {Jacobson}}]{Richman98OL}%
  \BibitemOpen
  \bibfield  {author} {\bibinfo {author} {\bibfnamefont {B.~A.}\ \bibnamefont
  {Richman}}, \bibinfo {author} {\bibfnamefont {S.~E.}\ \bibnamefont {Bisson}},
  \bibinfo {author} {\bibfnamefont {R.}~\bibnamefont {Trebino}}, \bibinfo
  {author} {\bibfnamefont {E.}~\bibnamefont {Sidick}},\ and\ \bibinfo {author}
  {\bibfnamefont {A.}~\bibnamefont {Jacobson}},\ }\bibfield  {title} {\bibinfo
  {title} {Efficient broadband second-harmonic generation by dispersive
  achromatic nonlinear conversion using only prisms},\ }\href@noop {}
  {\bibfield  {journal} {\bibinfo  {journal} {Opt. Lett}\ }\textbf {\bibinfo
  {volume} {23}},\ \bibinfo {pages} {497} (\bibinfo {year} {1998})}\BibitemShut
  {NoStop}%
\bibitem [{\citenamefont {Richman}\ \emph {et~al.}(1999)\citenamefont
  {Richman}, \citenamefont {Bisson}, \citenamefont {Trebino}, \citenamefont
  {Sidick},\ and\ \citenamefont {Jacobson}}]{Richman99AO}%
  \BibitemOpen
  \bibfield  {author} {\bibinfo {author} {\bibfnamefont {B.~A.}\ \bibnamefont
  {Richman}}, \bibinfo {author} {\bibfnamefont {S.~E.}\ \bibnamefont {Bisson}},
  \bibinfo {author} {\bibfnamefont {R.}~\bibnamefont {Trebino}}, \bibinfo
  {author} {\bibfnamefont {E.}~\bibnamefont {Sidick}},\ and\ \bibinfo {author}
  {\bibfnamefont {A.}~\bibnamefont {Jacobson}},\ }\bibfield  {title} {\bibinfo
  {title} {All-prism achromatic phase matching for tunable second-harmonic
  generation},\ }\href@noop {} {\bibfield  {journal} {\bibinfo  {journal}
  {Appl. Opt.}\ }\textbf {\bibinfo {volume} {38}},\ \bibinfo {pages} {3316}
  (\bibinfo {year} {1999})}\BibitemShut {NoStop}%
\bibitem [{\citenamefont {Nugraha}\ \emph {et~al.}(2019)\citenamefont
  {Nugraha}, \citenamefont {Krizs{\'a}n}, \citenamefont {Lombosi},
  \citenamefont {P{\'a}lfalvi}, \citenamefont {T{\'o}th}, \citenamefont
  {Alm{\'a}si}, \citenamefont {F{\:}l{\:o}p},\ and\ \citenamefont
  {Hebling}}]{Nugraha19OL}%
  \BibitemOpen
  \bibfield  {author} {\bibinfo {author} {\bibfnamefont {P.~S.}\ \bibnamefont
  {Nugraha}}, \bibinfo {author} {\bibfnamefont {G.}~\bibnamefont
  {Krizs{\'a}n}}, \bibinfo {author} {\bibfnamefont {C.}~\bibnamefont
  {Lombosi}}, \bibinfo {author} {\bibfnamefont {L.}~\bibnamefont
  {P{\'a}lfalvi}}, \bibinfo {author} {\bibfnamefont {G.}~\bibnamefont
  {T{\'o}th}}, \bibinfo {author} {\bibfnamefont {G.}~\bibnamefont
  {Alm{\'a}si}}, \bibinfo {author} {\bibfnamefont {J.~A.}\ \bibnamefont
  {F{\:}l{\:o}p}},\ and\ \bibinfo {author} {\bibfnamefont {J.}~\bibnamefont
  {Hebling}},\ }\bibfield  {title} {\bibinfo {title} {Demonstration of a
  tilted-pulse-front pumped plane-parallel slab terahertz source},\ }\href@noop
  {} {\bibfield  {journal} {\bibinfo  {journal} {Opt. Lett.}\ }\textbf
  {\bibinfo {volume} {44}},\ \bibinfo {pages} {1023} (\bibinfo {year}
  {2019})}\BibitemShut {NoStop}%
\bibitem [{\citenamefont {Wang}\ \emph {et~al.}(2020)\citenamefont {Wang},
  \citenamefont {T{\'o}th}, \citenamefont {Hebling},\ and\ \citenamefont
  {K{\"a}rtner}}]{Wang20LPR}%
  \BibitemOpen
  \bibfield  {author} {\bibinfo {author} {\bibfnamefont {L.}~\bibnamefont
  {Wang}}, \bibinfo {author} {\bibfnamefont {G.}~\bibnamefont {T{\'o}th}},
  \bibinfo {author} {\bibfnamefont {J.}~\bibnamefont {Hebling}},\ and\ \bibinfo
  {author} {\bibfnamefont {F.}~\bibnamefont {K{\"a}rtner}},\ }\bibfield
  {title} {\bibinfo {title} {Tilted-pulse-front schemes for terahertz
  generation},\ }\href@noop {} {\bibfield  {journal} {\bibinfo  {journal}
  {Laser Photon. Rev.}\ }\textbf {\bibinfo {volume} {14}},\ \bibinfo {pages}
  {2000021} (\bibinfo {year} {2020})}\BibitemShut {NoStop}%
\bibitem [{\citenamefont {Arbabi}\ \emph {et~al.}(2017)\citenamefont {Arbabi},
  \citenamefont {Arbabi}, \citenamefont {Kamali}, \citenamefont {Horie},\ and\
  \citenamefont {Faraon}}]{Arbabi17Optica}%
  \BibitemOpen
  \bibfield  {author} {\bibinfo {author} {\bibfnamefont {E.}~\bibnamefont
  {Arbabi}}, \bibinfo {author} {\bibfnamefont {A.}~\bibnamefont {Arbabi}},
  \bibinfo {author} {\bibfnamefont {S.~M.}\ \bibnamefont {Kamali}}, \bibinfo
  {author} {\bibfnamefont {Y.}~\bibnamefont {Horie}},\ and\ \bibinfo {author}
  {\bibfnamefont {A.}~\bibnamefont {Faraon}},\ }\bibfield  {title} {\bibinfo
  {title} {Controlling the sign of chromatic dispersion in diffractive optics
  with dielectric metasurfaces},\ }\href@noop {} {\bibfield  {journal}
  {\bibinfo  {journal} {Optica}\ }\textbf {\bibinfo {volume} {4}},\ \bibinfo
  {pages} {625} (\bibinfo {year} {2017})}\BibitemShut {NoStop}%
\bibitem [{\citenamefont {McClung}\ \emph {et~al.}(2020)\citenamefont
  {McClung}, \citenamefont {Mansouree},\ and\ \citenamefont
  {Arbabi}}]{McClung20Light}%
  \BibitemOpen
  \bibfield  {author} {\bibinfo {author} {\bibfnamefont {A.}~\bibnamefont
  {McClung}}, \bibinfo {author} {\bibfnamefont {M.}~\bibnamefont {Mansouree}},\
  and\ \bibinfo {author} {\bibfnamefont {A.}~\bibnamefont {Arbabi}},\
  }\bibfield  {title} {\bibinfo {title} {At-will chromatic dispersion by
  prescribing light trajectories with cascaded metasurfaces},\ }\href@noop {}
  {\bibfield  {journal} {\bibinfo  {journal} {Light Sci. Appl.}\ }\textbf
  {\bibinfo {volume} {9}},\ \bibinfo {pages} {93} (\bibinfo {year}
  {2020})}\BibitemShut {NoStop}%
\bibitem [{\citenamefont {Shaltout}\ \emph {et~al.}(2019)\citenamefont
  {Shaltout}, \citenamefont {Lagoudakis}, \citenamefont {{van de G}roep},
  \citenamefont {Kim}, \citenamefont {Vu\v{c}kovi{\'c}}, \citenamefont
  {Shalaev},\ and\ \citenamefont {Brongersma}}]{Shaltout19ScienceSteering}%
  \BibitemOpen
  \bibfield  {author} {\bibinfo {author} {\bibfnamefont {A.~M.}\ \bibnamefont
  {Shaltout}}, \bibinfo {author} {\bibfnamefont {K.~G.}\ \bibnamefont
  {Lagoudakis}}, \bibinfo {author} {\bibfnamefont {J.}~\bibnamefont {{van de
  G}roep}}, \bibinfo {author} {\bibfnamefont {S.~J.}\ \bibnamefont {Kim}},
  \bibinfo {author} {\bibfnamefont {J.}~\bibnamefont {Vu\v{c}kovi{\'c}}},
  \bibinfo {author} {\bibfnamefont {V.~M.}\ \bibnamefont {Shalaev}},\ and\
  \bibinfo {author} {\bibfnamefont {M.~L.}\ \bibnamefont {Brongersma}},\
  }\bibfield  {title} {\bibinfo {title} {Spatiotemporal light control with
  frequency-gradient metasurfaces},\ }\href@noop {} {\bibfield  {journal}
  {\bibinfo  {journal} {Science}\ }\textbf {\bibinfo {volume} {365}},\ \bibinfo
  {pages} {374} (\bibinfo {year} {2019})}\BibitemShut {NoStop}%
\bibitem [{\citenamefont {Porras}\ \emph {et~al.}(2003)\citenamefont {Porras},
  \citenamefont {Valiulis},\ and\ \citenamefont {{Di T}rapani}}]{Porras03PRE2}%
  \BibitemOpen
  \bibfield  {author} {\bibinfo {author} {\bibfnamefont {M.~A.}\ \bibnamefont
  {Porras}}, \bibinfo {author} {\bibfnamefont {G.}~\bibnamefont {Valiulis}},\
  and\ \bibinfo {author} {\bibfnamefont {P.}~\bibnamefont {{Di T}rapani}},\
  }\bibfield  {title} {\bibinfo {title} {Unified description of {B}essel {X}
  waves with cone dispersion and tilted pulses},\ }\href@noop {} {\bibfield
  {journal} {\bibinfo  {journal} {Phys. Rev. E}\ }\textbf {\bibinfo {volume}
  {68}},\ \bibinfo {pages} {016613} (\bibinfo {year} {2003})}\BibitemShut
  {NoStop}%
\bibitem [{\citenamefont {Hall}\ \emph
  {et~al.}(2021{\natexlab{a}})\citenamefont {Hall}, \citenamefont {Yessenov},\
  and\ \citenamefont {Abouraddy}}]{Hall21OL}%
  \BibitemOpen
  \bibfield  {author} {\bibinfo {author} {\bibfnamefont {L.~A.}\ \bibnamefont
  {Hall}}, \bibinfo {author} {\bibfnamefont {M.}~\bibnamefont {Yessenov}},\
  and\ \bibinfo {author} {\bibfnamefont {A.~F.}\ \bibnamefont {Abouraddy}},\
  }\bibfield  {title} {\bibinfo {title} {Space-time wave packets violate the
  universal relationship between angular dispersion and pulse-front tilt},\
  }\href@noop {} {\bibfield  {journal} {\bibinfo  {journal} {Opt. Lett.}\
  }\textbf {\bibinfo {volume} {46}},\ \bibinfo {pages} {1672} (\bibinfo {year}
  {2021}{\natexlab{a}})}\BibitemShut {NoStop}%
\bibitem [{\citenamefont {Yessenov}\ \emph {et~al.}(2021)\citenamefont
  {Yessenov}, \citenamefont {Hall},\ and\ \citenamefont
  {Abouraddy}}]{Yessenov21ACSP}%
  \BibitemOpen
  \bibfield  {author} {\bibinfo {author} {\bibfnamefont {M.}~\bibnamefont
  {Yessenov}}, \bibinfo {author} {\bibfnamefont {L.~A.}\ \bibnamefont {Hall}},\
  and\ \bibinfo {author} {\bibfnamefont {A.~F.}\ \bibnamefont {Abouraddy}},\
  }\bibfield  {title} {\bibinfo {title} {Engineering the optical vacuum:
  {A}rbitrary magnitude, sign, and order of dispersion in free space using
  space-time wave packets},\ }\href@noop {} {\bibfield  {journal} {\bibinfo
  {journal} {ACS Photon.}\ }\textbf {\bibinfo {volume} {8}},\ \bibinfo {pages}
  {2274} (\bibinfo {year} {2021})}\BibitemShut {NoStop}%
\bibitem [{\citenamefont {Hall}\ and\ \citenamefont
  {Abouraddy}(2021{\natexlab{a}})}]{Hall21OL3NormalGVD}%
  \BibitemOpen
  \bibfield  {author} {\bibinfo {author} {\bibfnamefont {L.~A.}\ \bibnamefont
  {Hall}}\ and\ \bibinfo {author} {\bibfnamefont {A.~F.}\ \bibnamefont
  {Abouraddy}},\ }\bibfield  {title} {\bibinfo {title} {Realizing normal
  group-velocity dispersion in free space via angular dispersion},\ }\href@noop
  {} {\bibfield  {journal} {\bibinfo  {journal} {arXiv:2108.00312}\ } (\bibinfo
  {year} {2021}{\natexlab{a}})}\BibitemShut {NoStop}%
\bibitem [{\citenamefont {Hall}\ and\ \citenamefont
  {Abouraddy}(2021{\natexlab{b}})}]{Hall21OE1NonDiff}%
  \BibitemOpen
  \bibfield  {author} {\bibinfo {author} {\bibfnamefont {L.~A.}\ \bibnamefont
  {Hall}}\ and\ \bibinfo {author} {\bibfnamefont {A.~F.}\ \bibnamefont
  {Abouraddy}},\ }\bibfield  {title} {\bibinfo {title} {The consequences of
  non-differentiable angular dispersion in optics: {T}ilted pulse fronts versus
  space-time wave packets},\ }\href@noop {} {\bibfield  {journal} {\bibinfo
  {journal} {arXiv:2109.07039}\ } (\bibinfo {year}
  {2021}{\natexlab{b}})}\BibitemShut {NoStop}%
\bibitem [{\citenamefont {Kondakci}\ and\ \citenamefont
  {Abouraddy}(2016)}]{Kondakci16OE}%
  \BibitemOpen
  \bibfield  {author} {\bibinfo {author} {\bibfnamefont {H.~E.}\ \bibnamefont
  {Kondakci}}\ and\ \bibinfo {author} {\bibfnamefont {A.~F.}\ \bibnamefont
  {Abouraddy}},\ }\bibfield  {title} {\bibinfo {title} {Diffraction-free pulsed
  optical beams via space-time correlations},\ }\href@noop {} {\bibfield
  {journal} {\bibinfo  {journal} {Opt. Express}\ }\textbf {\bibinfo {volume}
  {24}},\ \bibinfo {pages} {28659} (\bibinfo {year} {2016})}\BibitemShut
  {NoStop}%
\bibitem [{\citenamefont {Parker}\ and\ \citenamefont
  {Alonso}(2016)}]{Parker16OE}%
  \BibitemOpen
  \bibfield  {author} {\bibinfo {author} {\bibfnamefont {K.~J.}\ \bibnamefont
  {Parker}}\ and\ \bibinfo {author} {\bibfnamefont {M.~A.}\ \bibnamefont
  {Alonso}},\ }\bibfield  {title} {\bibinfo {title} {The longitudinal iso-phase
  condition and needle pulses},\ }\href@noop {} {\bibfield  {journal} {\bibinfo
   {journal} {Opt. Express}\ }\textbf {\bibinfo {volume} {24}},\ \bibinfo
  {pages} {28669} (\bibinfo {year} {2016})}\BibitemShut {NoStop}%
\bibitem [{\citenamefont {Kondakci}\ and\ \citenamefont
  {Abouraddy}(2017)}]{Kondakci17NP}%
  \BibitemOpen
  \bibfield  {author} {\bibinfo {author} {\bibfnamefont {H.~E.}\ \bibnamefont
  {Kondakci}}\ and\ \bibinfo {author} {\bibfnamefont {A.~F.}\ \bibnamefont
  {Abouraddy}},\ }\bibfield  {title} {\bibinfo {title} {Diffraction-free
  space-time beams},\ }\href@noop {} {\bibfield  {journal} {\bibinfo  {journal}
  {Nat. Photon.}\ }\textbf {\bibinfo {volume} {11}},\ \bibinfo {pages} {733}
  (\bibinfo {year} {2017})}\BibitemShut {NoStop}%
\bibitem [{\citenamefont {Porras}(2017)}]{Porras17OL}%
  \BibitemOpen
  \bibfield  {author} {\bibinfo {author} {\bibfnamefont {M.~A.}\ \bibnamefont
  {Porras}},\ }\bibfield  {title} {\bibinfo {title} {Gaussian beams diffracting
  in time},\ }\href@noop {} {\bibfield  {journal} {\bibinfo  {journal} {Opt.
  Lett.}\ }\textbf {\bibinfo {volume} {42}},\ \bibinfo {pages} {4679} (\bibinfo
  {year} {2017})}\BibitemShut {NoStop}%
\bibitem [{\citenamefont {Efremidis}(2017)}]{Efremidis17OL}%
  \BibitemOpen
  \bibfield  {author} {\bibinfo {author} {\bibfnamefont {N.~K.}\ \bibnamefont
  {Efremidis}},\ }\bibfield  {title} {\bibinfo {title} {Spatiotemporal
  diffraction-free pulsed beams in free-space of the {A}iry and {B}essel
  type},\ }\href@noop {} {\bibfield  {journal} {\bibinfo  {journal} {Opt.
  Lett.}\ }\textbf {\bibinfo {volume} {42}},\ \bibinfo {pages} {5038} (\bibinfo
  {year} {2017})}\BibitemShut {NoStop}%
\bibitem [{\citenamefont {Yessenov}\ \emph
  {et~al.}(2019{\natexlab{a}})\citenamefont {Yessenov}, \citenamefont
  {Bhaduri}, \citenamefont {Kondakci},\ and\ \citenamefont
  {Abouraddy}}]{Yessenov19OPN}%
  \BibitemOpen
  \bibfield  {author} {\bibinfo {author} {\bibfnamefont {M.}~\bibnamefont
  {Yessenov}}, \bibinfo {author} {\bibfnamefont {B.}~\bibnamefont {Bhaduri}},
  \bibinfo {author} {\bibfnamefont {H.~E.}\ \bibnamefont {Kondakci}},\ and\
  \bibinfo {author} {\bibfnamefont {A.~F.}\ \bibnamefont {Abouraddy}},\
  }\bibfield  {title} {\bibinfo {title} {Weaving the rainbow: Space-time
  optical wave packets},\ }\href@noop {} {\bibfield  {journal} {\bibinfo
  {journal} {Opt. Photon. News}\ }\textbf {\bibinfo {volume} {30}},\ \bibinfo
  {pages} {34} (\bibinfo {year} {2019}{\natexlab{a}})}\BibitemShut {NoStop}%
\bibitem [{\citenamefont {Wong}(2021)}]{Wong21OE}%
  \BibitemOpen
  \bibfield  {author} {\bibinfo {author} {\bibfnamefont {L.~J.}\ \bibnamefont
  {Wong}},\ }\bibfield  {title} {\bibinfo {title} {Propagation-invariant
  space-time caustics of light},\ }\href@noop {} {\bibfield  {journal}
  {\bibinfo  {journal} {Opt. Express}\ }\textbf {\bibinfo {volume} {29}},\
  \bibinfo {pages} {30682} (\bibinfo {year} {2021})}\BibitemShut {NoStop}%
\bibitem [{\citenamefont {Kondakci}\ and\ \citenamefont
  {Abouraddy}(2018{\natexlab{a}})}]{Kondakci18PRL}%
  \BibitemOpen
  \bibfield  {author} {\bibinfo {author} {\bibfnamefont {H.~E.}\ \bibnamefont
  {Kondakci}}\ and\ \bibinfo {author} {\bibfnamefont {A.~F.}\ \bibnamefont
  {Abouraddy}},\ }\bibfield  {title} {\bibinfo {title} {Airy wavepackets
  accelerating in space-time},\ }\href@noop {} {\bibfield  {journal} {\bibinfo
  {journal} {Phys. Rev. Lett.}\ }\textbf {\bibinfo {volume} {120}},\ \bibinfo
  {pages} {163901} (\bibinfo {year} {2018}{\natexlab{a}})}\BibitemShut
  {NoStop}%
\bibitem [{\citenamefont {Bhaduri}\ \emph {et~al.}(2019)\citenamefont
  {Bhaduri}, \citenamefont {Yessenov}, \citenamefont {Reyes}, \citenamefont
  {Pena}, \citenamefont {Meem}, \citenamefont {Fairchild}, \citenamefont
  {Menon}, \citenamefont {Richardson},\ and\ \citenamefont
  {Abouraddy}}]{Bhaduri19OL}%
  \BibitemOpen
  \bibfield  {author} {\bibinfo {author} {\bibfnamefont {B.}~\bibnamefont
  {Bhaduri}}, \bibinfo {author} {\bibfnamefont {M.}~\bibnamefont {Yessenov}},
  \bibinfo {author} {\bibfnamefont {D.}~\bibnamefont {Reyes}}, \bibinfo
  {author} {\bibfnamefont {J.}~\bibnamefont {Pena}}, \bibinfo {author}
  {\bibfnamefont {M.}~\bibnamefont {Meem}}, \bibinfo {author} {\bibfnamefont
  {S.~R.}\ \bibnamefont {Fairchild}}, \bibinfo {author} {\bibfnamefont
  {R.}~\bibnamefont {Menon}}, \bibinfo {author} {\bibfnamefont {M.~C.}\
  \bibnamefont {Richardson}},\ and\ \bibinfo {author} {\bibfnamefont {A.~F.}\
  \bibnamefont {Abouraddy}},\ }\bibfield  {title} {\bibinfo {title} {Broadband
  space-time wave packets propagating 70~m},\ }\href@noop {} {\bibfield
  {journal} {\bibinfo  {journal} {Opt. Lett.}\ }\textbf {\bibinfo {volume}
  {44}},\ \bibinfo {pages} {2073} (\bibinfo {year} {2019})}\BibitemShut
  {NoStop}%
\bibitem [{\citenamefont {Yessenov}\ \emph
  {et~al.}(2019{\natexlab{b}})\citenamefont {Yessenov}, \citenamefont
  {Bhaduri}, \citenamefont {Mach}, \citenamefont {Mardani}, \citenamefont
  {Kondakci}, \citenamefont {Alonso}, \citenamefont {Atia},\ and\ \citenamefont
  {Abouraddy}}]{Yessenov19OE}%
  \BibitemOpen
  \bibfield  {author} {\bibinfo {author} {\bibfnamefont {M.}~\bibnamefont
  {Yessenov}}, \bibinfo {author} {\bibfnamefont {B.}~\bibnamefont {Bhaduri}},
  \bibinfo {author} {\bibfnamefont {L.}~\bibnamefont {Mach}}, \bibinfo {author}
  {\bibfnamefont {D.}~\bibnamefont {Mardani}}, \bibinfo {author} {\bibfnamefont
  {H.~E.}\ \bibnamefont {Kondakci}}, \bibinfo {author} {\bibfnamefont {M.~A.}\
  \bibnamefont {Alonso}}, \bibinfo {author} {\bibfnamefont {G.~A.}\
  \bibnamefont {Atia}},\ and\ \bibinfo {author} {\bibfnamefont {A.~F.}\
  \bibnamefont {Abouraddy}},\ }\bibfield  {title} {\bibinfo {title} {What is
  the maximum differential group delay achievable by a space-time wave packet
  in free space?},\ }\href@noop {} {\bibfield  {journal} {\bibinfo  {journal}
  {Opt. Express}\ }\textbf {\bibinfo {volume} {27}},\ \bibinfo {pages} {12443}
  (\bibinfo {year} {2019}{\natexlab{b}})}\BibitemShut {NoStop}%
\bibitem [{\citenamefont {Yessenov}\ \emph
  {et~al.}(2019{\natexlab{c}})\citenamefont {Yessenov}, \citenamefont
  {Bhaduri}, \citenamefont {Kondakci}, \citenamefont {Meem}, \citenamefont
  {Menon},\ and\ \citenamefont {Abouraddy}}]{Yessenov19Optica}%
  \BibitemOpen
  \bibfield  {author} {\bibinfo {author} {\bibfnamefont {M.}~\bibnamefont
  {Yessenov}}, \bibinfo {author} {\bibfnamefont {B.}~\bibnamefont {Bhaduri}},
  \bibinfo {author} {\bibfnamefont {H.~E.}\ \bibnamefont {Kondakci}}, \bibinfo
  {author} {\bibfnamefont {M.}~\bibnamefont {Meem}}, \bibinfo {author}
  {\bibfnamefont {R.}~\bibnamefont {Menon}},\ and\ \bibinfo {author}
  {\bibfnamefont {A.~F.}\ \bibnamefont {Abouraddy}},\ }\bibfield  {title}
  {\bibinfo {title} {Non-diffracting broadband incoherent space–time
  fields},\ }\href@noop {} {\bibfield  {journal} {\bibinfo  {journal} {Optica}\
  }\textbf {\bibinfo {volume} {6}},\ \bibinfo {pages} {522} (\bibinfo {year}
  {2019}{\natexlab{c}})}\BibitemShut {NoStop}%
\bibitem [{\citenamefont {Yessenov}\ \emph
  {et~al.}(2020{\natexlab{a}})\citenamefont {Yessenov}, \citenamefont
  {Bhaduri}, \citenamefont {Delfyett},\ and\ \citenamefont
  {Abouraddy}}]{Yessenov20NC}%
  \BibitemOpen
  \bibfield  {author} {\bibinfo {author} {\bibfnamefont {M.}~\bibnamefont
  {Yessenov}}, \bibinfo {author} {\bibfnamefont {B.}~\bibnamefont {Bhaduri}},
  \bibinfo {author} {\bibfnamefont {P.~J.}\ \bibnamefont {Delfyett}},\ and\
  \bibinfo {author} {\bibfnamefont {A.~F.}\ \bibnamefont {Abouraddy}},\
  }\bibfield  {title} {\bibinfo {title} {Free-space optical delay line using
  space-time wave packets},\ }\href@noop {} {\bibfield  {journal} {\bibinfo
  {journal} {Nat. Commun.}\ }\textbf {\bibinfo {volume} {11}},\ \bibinfo
  {pages} {5782} (\bibinfo {year} {2020}{\natexlab{a}})}\BibitemShut {NoStop}%
\bibitem [{\citenamefont {Schepler}\ \emph {et~al.}(2020)\citenamefont
  {Schepler}, \citenamefont {Yessenov}, \citenamefont {Zhiyenbayev},\ and\
  \citenamefont {Abouraddy}}]{Schepler20ACSP}%
  \BibitemOpen
  \bibfield  {author} {\bibinfo {author} {\bibfnamefont {K.~L.}\ \bibnamefont
  {Schepler}}, \bibinfo {author} {\bibfnamefont {M.}~\bibnamefont {Yessenov}},
  \bibinfo {author} {\bibfnamefont {Y.}~\bibnamefont {Zhiyenbayev}},\ and\
  \bibinfo {author} {\bibfnamefont {A.~F.}\ \bibnamefont {Abouraddy}},\
  }\bibfield  {title} {\bibinfo {title} {Space--time surface plasmon
  polaritons: A new propagation-invariant surface wave packet},\ }\href@noop {}
  {\bibfield  {journal} {\bibinfo  {journal} {ACS Photon.}\ }\textbf {\bibinfo
  {volume} {7}},\ \bibinfo {pages} {2966} (\bibinfo {year} {2020})}\BibitemShut
  {NoStop}%
\bibitem [{\citenamefont {Wong}\ \emph {et~al.}(2020)\citenamefont {Wong},
  \citenamefont {Christodoulides},\ and\ \citenamefont {Kaminer}}]{Wong20AS}%
  \BibitemOpen
  \bibfield  {author} {\bibinfo {author} {\bibfnamefont {L.~J.}\ \bibnamefont
  {Wong}}, \bibinfo {author} {\bibfnamefont {D.~N.}\ \bibnamefont
  {Christodoulides}},\ and\ \bibinfo {author} {\bibfnamefont {I.}~\bibnamefont
  {Kaminer}},\ }\bibfield  {title} {\bibinfo {title} {The complex charge
  paradigm: {A} new approach for designing electromagnetic wavepackets},\
  }\href@noop {} {\bibfield  {journal} {\bibinfo  {journal} {Adv. Sci.}\
  }\textbf {\bibinfo {volume} {7}},\ \bibinfo {pages} {1903377} (\bibinfo
  {year} {2020})}\BibitemShut {NoStop}%
\bibitem [{\citenamefont {Wong}\ and\ \citenamefont
  {Kaminer}(2017)}]{Wong17ACSP2}%
  \BibitemOpen
  \bibfield  {author} {\bibinfo {author} {\bibfnamefont {L.~J.}\ \bibnamefont
  {Wong}}\ and\ \bibinfo {author} {\bibfnamefont {I.}~\bibnamefont {Kaminer}},\
  }\bibfield  {title} {\bibinfo {title} {Ultrashort tilted-pulsefront pulses
  and nonparaxial tilted-phase-front beams},\ }\href@noop {} {\bibfield
  {journal} {\bibinfo  {journal} {ACS Photon.}\ }\textbf {\bibinfo {volume}
  {4}},\ \bibinfo {pages} {2257} (\bibinfo {year} {2017})}\BibitemShut
  {NoStop}%
\bibitem [{\citenamefont {Kondakci}\ and\ \citenamefont
  {Abouraddy}(2019)}]{Kondakci19NC}%
  \BibitemOpen
  \bibfield  {author} {\bibinfo {author} {\bibfnamefont {H.~E.}\ \bibnamefont
  {Kondakci}}\ and\ \bibinfo {author} {\bibfnamefont {A.~F.}\ \bibnamefont
  {Abouraddy}},\ }\bibfield  {title} {\bibinfo {title} {Optical space-time wave
  packets of arbitrary group velocity in free space},\ }\href@noop {}
  {\bibfield  {journal} {\bibinfo  {journal} {Nat. Commun.}\ }\textbf {\bibinfo
  {volume} {10}},\ \bibinfo {pages} {929} (\bibinfo {year} {2019})}\BibitemShut
  {NoStop}%
\bibitem [{\citenamefont {Kondakci}\ and\ \citenamefont
  {Abouraddy}(2018{\natexlab{b}})}]{Kondakci18OL}%
  \BibitemOpen
  \bibfield  {author} {\bibinfo {author} {\bibfnamefont {H.~E.}\ \bibnamefont
  {Kondakci}}\ and\ \bibinfo {author} {\bibfnamefont {A.~F.}\ \bibnamefont
  {Abouraddy}},\ }\bibfield  {title} {\bibinfo {title} {Self-healing of
  space-time light sheets},\ }\href@noop {} {\bibfield  {journal} {\bibinfo
  {journal} {Opt. Lett.}\ }\textbf {\bibinfo {volume} {43}},\ \bibinfo {pages}
  {3830} (\bibinfo {year} {2018}{\natexlab{b}})}\BibitemShut {NoStop}%
\bibitem [{\citenamefont {Hall}\ \emph
  {et~al.}(2021{\natexlab{b}})\citenamefont {Hall}, \citenamefont {Yessenov},
  \citenamefont {Ponomarenko},\ and\ \citenamefont {Abouraddy}}]{Hall21APLP}%
  \BibitemOpen
  \bibfield  {author} {\bibinfo {author} {\bibfnamefont {L.~A.}\ \bibnamefont
  {Hall}}, \bibinfo {author} {\bibfnamefont {M.}~\bibnamefont {Yessenov}},
  \bibinfo {author} {\bibfnamefont {S.~A.}\ \bibnamefont {Ponomarenko}},\ and\
  \bibinfo {author} {\bibfnamefont {A.~F.}\ \bibnamefont {Abouraddy}},\
  }\bibfield  {title} {\bibinfo {title} {The space-time {T}albot effect},\
  }\href@noop {} {\bibfield  {journal} {\bibinfo  {journal} {APL Photon.}\
  }\textbf {\bibinfo {volume} {6}},\ \bibinfo {pages} {056105} (\bibinfo {year}
  {2021}{\natexlab{b}})}\BibitemShut {NoStop}%
\bibitem [{\citenamefont {Hall}\ \emph
  {et~al.}(2021{\natexlab{c}})\citenamefont {Hall}, \citenamefont
  {Ponomarenko},\ and\ \citenamefont {Abouraddy}}]{Hall21OL2}%
  \BibitemOpen
  \bibfield  {author} {\bibinfo {author} {\bibfnamefont {L.~A.}\ \bibnamefont
  {Hall}}, \bibinfo {author} {\bibfnamefont {S.}~\bibnamefont {Ponomarenko}},\
  and\ \bibinfo {author} {\bibfnamefont {A.~F.}\ \bibnamefont {Abouraddy}},\
  }\bibfield  {title} {\bibinfo {title} {Temporal {T}albot effect in free
  space},\ }\href@noop {} {\bibfield  {journal} {\bibinfo  {journal} {Opt.
  Lett.}\ }\textbf {\bibinfo {volume} {46}},\ \bibinfo {pages} {3107} (\bibinfo
  {year} {2021}{\natexlab{c}})}\BibitemShut {NoStop}%
\bibitem [{\citenamefont {Yessenov}\ and\ \citenamefont
  {Abouraddy}(2020)}]{Yessenov20PRL2}%
  \BibitemOpen
  \bibfield  {author} {\bibinfo {author} {\bibfnamefont {M.}~\bibnamefont
  {Yessenov}}\ and\ \bibinfo {author} {\bibfnamefont {A.~F.}\ \bibnamefont
  {Abouraddy}},\ }\bibfield  {title} {\bibinfo {title} {Accelerating and
  decelerating space-time wave packets in free space},\ }\href@noop {}
  {\bibfield  {journal} {\bibinfo  {journal} {Phys. Rev. Lett.}\ }\textbf
  {\bibinfo {volume} {125}},\ \bibinfo {pages} {233901} (\bibinfo {year}
  {2020})}\BibitemShut {NoStop}%
\bibitem [{\citenamefont {Hall}\ \emph
  {et~al.}(2021{\natexlab{d}})\citenamefont {Hall}, \citenamefont {Yessenov},\
  and\ \citenamefont {Abouraddy}}]{Hall21OL4Acceleration}%
  \BibitemOpen
  \bibfield  {author} {\bibinfo {author} {\bibfnamefont {L.~A.}\ \bibnamefont
  {Hall}}, \bibinfo {author} {\bibfnamefont {M.}~\bibnamefont {Yessenov}},\
  and\ \bibinfo {author} {\bibfnamefont {A.~F.}\ \bibnamefont {Abouraddy}},\
  }\bibfield  {title} {\bibinfo {title} {Arbitrarily accelerating space-time
  wave packets},\ }\href@noop {} {\bibfield  {journal} {\bibinfo  {journal}
  {arXiv:2109.04009}\ } (\bibinfo {year} {2021}{\natexlab{d}})}\BibitemShut
  {NoStop}%
\bibitem [{\citenamefont {Malaguti}\ \emph {et~al.}(2008)\citenamefont
  {Malaguti}, \citenamefont {Bellanca},\ and\ \citenamefont
  {Trillo}}]{Malaguti08OL}%
  \BibitemOpen
  \bibfield  {author} {\bibinfo {author} {\bibfnamefont {S.}~\bibnamefont
  {Malaguti}}, \bibinfo {author} {\bibfnamefont {G.}~\bibnamefont {Bellanca}},\
  and\ \bibinfo {author} {\bibfnamefont {S.}~\bibnamefont {Trillo}},\
  }\bibfield  {title} {\bibinfo {title} {Two-dimensional envelope localized
  waves in the anomalous dispersion regime},\ }\href@noop {} {\bibfield
  {journal} {\bibinfo  {journal} {Opt. Lett.}\ }\textbf {\bibinfo {volume}
  {33}},\ \bibinfo {pages} {1117} (\bibinfo {year} {2008})}\BibitemShut
  {NoStop}%
\bibitem [{\citenamefont {Malaguti}\ and\ \citenamefont
  {Trillo}(2009)}]{Malaguti09PRA}%
  \BibitemOpen
  \bibfield  {author} {\bibinfo {author} {\bibfnamefont {S.}~\bibnamefont
  {Malaguti}}\ and\ \bibinfo {author} {\bibfnamefont {S.}~\bibnamefont
  {Trillo}},\ }\bibfield  {title} {\bibinfo {title} {Envelope localized waves
  of the conical type in linear normally dispersive media},\ }\href@noop {}
  {\bibfield  {journal} {\bibinfo  {journal} {Phys. Rev. A}\ }\textbf {\bibinfo
  {volume} {79}},\ \bibinfo {pages} {063803} (\bibinfo {year}
  {2009})}\BibitemShut {NoStop}%
\bibitem [{\citenamefont {Bhaduri}\ \emph {et~al.}(2020)\citenamefont
  {Bhaduri}, \citenamefont {Yessenov},\ and\ \citenamefont
  {Abouraddy}}]{Bhaduri20NP}%
  \BibitemOpen
  \bibfield  {author} {\bibinfo {author} {\bibfnamefont {B.}~\bibnamefont
  {Bhaduri}}, \bibinfo {author} {\bibfnamefont {M.}~\bibnamefont {Yessenov}},\
  and\ \bibinfo {author} {\bibfnamefont {A.~F.}\ \bibnamefont {Abouraddy}},\
  }\bibfield  {title} {\bibinfo {title} {Anomalous refraction of optical
  spacetime wave packets},\ }\href@noop {} {\bibfield  {journal} {\bibinfo
  {journal} {Nat. Photon.}\ }\textbf {\bibinfo {volume} {14}},\ \bibinfo
  {pages} {416} (\bibinfo {year} {2020})}\BibitemShut {NoStop}%
\bibitem [{\citenamefont {Chiao}\ and\ \citenamefont
  {Milonni}(2002)}]{Chiao02OPN}%
  \BibitemOpen
  \bibfield  {author} {\bibinfo {author} {\bibfnamefont {R.~Y.}\ \bibnamefont
  {Chiao}}\ and\ \bibinfo {author} {\bibfnamefont {P.~W.}\ \bibnamefont
  {Milonni}},\ }\bibfield  {title} {\bibinfo {title} {Fast light, slow light},\
  }\href@noop {} {\bibfield  {journal} {\bibinfo  {journal} {Opt. Photon.
  News}\ }\textbf {\bibinfo {volume} {13}},\ \bibinfo {pages} {26} (\bibinfo
  {year} {2002})}\BibitemShut {NoStop}%
\bibitem [{\citenamefont {Hebling}(1996)}]{Hebling96OQE}%
  \BibitemOpen
  \bibfield  {author} {\bibinfo {author} {\bibfnamefont {J.}~\bibnamefont
  {Hebling}},\ }\bibfield  {title} {\bibinfo {title} {Derivation of the pulse
  front tilt caused by angular dispersion},\ }\href@noop {} {\bibfield
  {journal} {\bibinfo  {journal} {Opt. Quant. Electron.}\ }\textbf {\bibinfo
  {volume} {28}},\ \bibinfo {pages} {1759} (\bibinfo {year}
  {1996})}\BibitemShut {NoStop}%
\bibitem [{\citenamefont {Yessenov}\ \emph
  {et~al.}(2019{\natexlab{d}})\citenamefont {Yessenov}, \citenamefont
  {Bhaduri}, \citenamefont {Kondakci},\ and\ \citenamefont
  {Abouraddy}}]{Yessenov19PRA}%
  \BibitemOpen
  \bibfield  {author} {\bibinfo {author} {\bibfnamefont {M.}~\bibnamefont
  {Yessenov}}, \bibinfo {author} {\bibfnamefont {B.}~\bibnamefont {Bhaduri}},
  \bibinfo {author} {\bibfnamefont {H.~E.}\ \bibnamefont {Kondakci}},\ and\
  \bibinfo {author} {\bibfnamefont {A.~F.}\ \bibnamefont {Abouraddy}},\
  }\bibfield  {title} {\bibinfo {title} {Classification of
  propagation-invariant space-time light-sheets in free space: Theory and
  experiments},\ }\href@noop {} {\bibfield  {journal} {\bibinfo  {journal}
  {Phys. Rev. A}\ }\textbf {\bibinfo {volume} {99}},\ \bibinfo {pages} {023856}
  (\bibinfo {year} {2019}{\natexlab{d}})}\BibitemShut {NoStop}%
\bibitem [{\citenamefont {Donnelly}\ and\ \citenamefont
  {Ziolkowski}(1993)}]{Donnelly93ProcRSLA}%
  \BibitemOpen
  \bibfield  {author} {\bibinfo {author} {\bibfnamefont {R.}~\bibnamefont
  {Donnelly}}\ and\ \bibinfo {author} {\bibfnamefont {R.~W.}\ \bibnamefont
  {Ziolkowski}},\ }\bibfield  {title} {\bibinfo {title} {Designing localized
  waves},\ }\href@noop {} {\bibfield  {journal} {\bibinfo  {journal} {Proc. R.
  Soc. Lond. A}\ }\textbf {\bibinfo {volume} {440}},\ \bibinfo {pages} {541}
  (\bibinfo {year} {1993})}\BibitemShut {NoStop}%
\bibitem [{\citenamefont {Valtna}\ \emph {et~al.}(2007)\citenamefont {Valtna},
  \citenamefont {Reivelt},\ and\ \citenamefont {Saari}}]{Valtna07OC}%
  \BibitemOpen
  \bibfield  {author} {\bibinfo {author} {\bibfnamefont {H.}~\bibnamefont
  {Valtna}}, \bibinfo {author} {\bibfnamefont {K.}~\bibnamefont {Reivelt}},\
  and\ \bibinfo {author} {\bibfnamefont {P.}~\bibnamefont {Saari}},\ }\bibfield
   {title} {\bibinfo {title} {Methods for generating wideband localized waves
  of superluminal group velocity},\ }\href@noop {} {\bibfield  {journal}
  {\bibinfo  {journal} {Opt. Commun.}\ }\textbf {\bibinfo {volume} {278}},\
  \bibinfo {pages} {1} (\bibinfo {year} {2007})}\BibitemShut {NoStop}%
\bibitem [{\citenamefont {Zamboni-Rached}(2009)}]{ZamboniRached2009PRA}%
  \BibitemOpen
  \bibfield  {author} {\bibinfo {author} {\bibfnamefont {M.}~\bibnamefont
  {Zamboni-Rached}},\ }\bibfield  {title} {\bibinfo {title} {Unidirectional
  decomposition method for obtaining exact localized wave solutions totally
  free of backward components},\ }\href@noop {} {\bibfield  {journal} {\bibinfo
   {journal} {Phys. Rev. A}\ }\textbf {\bibinfo {volume} {79}},\ \bibinfo
  {pages} {013816} (\bibinfo {year} {2009})}\BibitemShut {NoStop}%
\bibitem [{\citenamefont {Brittingham}(1983)}]{Brittingham83JAP}%
  \BibitemOpen
  \bibfield  {author} {\bibinfo {author} {\bibfnamefont {J.~N.}\ \bibnamefont
  {Brittingham}},\ }\bibfield  {title} {\bibinfo {title} {Focus wave modes in
  homogeneous {M}axwell's equations: {T}ransverse electric mode},\ }\href@noop
  {} {\bibfield  {journal} {\bibinfo  {journal} {J. Appl. Phys.}\ }\textbf
  {\bibinfo {volume} {54}},\ \bibinfo {pages} {1179} (\bibinfo {year}
  {1983})}\BibitemShut {NoStop}%
\bibitem [{\citenamefont {Lu}\ and\ \citenamefont
  {Greenleaf}(1992)}]{Lu92IEEEa}%
  \BibitemOpen
  \bibfield  {author} {\bibinfo {author} {\bibfnamefont {J.-Y.}\ \bibnamefont
  {Lu}}\ and\ \bibinfo {author} {\bibfnamefont {J.~F.}\ \bibnamefont
  {Greenleaf}},\ }\bibfield  {title} {\bibinfo {title} {Nondiffracting {X}
  waves -- exact solutions to free-space scalar wave equation and their finite
  aperture realizations},\ }\href@noop {} {\bibfield  {journal} {\bibinfo
  {journal} {IEEE Trans. Ultrason. Ferroelec. Freq. Control}\ }\textbf
  {\bibinfo {volume} {39}},\ \bibinfo {pages} {19} (\bibinfo {year}
  {1992})}\BibitemShut {NoStop}%
\bibitem [{\citenamefont {Saari}\ and\ \citenamefont
  {Reivelt}(1997)}]{Saari97PRL}%
  \BibitemOpen
  \bibfield  {author} {\bibinfo {author} {\bibfnamefont {P.}~\bibnamefont
  {Saari}}\ and\ \bibinfo {author} {\bibfnamefont {K.}~\bibnamefont
  {Reivelt}},\ }\bibfield  {title} {\bibinfo {title} {Evidence of {X}-shaped
  propagation-invariant localized light waves},\ }\href@noop {} {\bibfield
  {journal} {\bibinfo  {journal} {Phys. Rev. Lett.}\ }\textbf {\bibinfo
  {volume} {79}},\ \bibinfo {pages} {4135} (\bibinfo {year}
  {1997})}\BibitemShut {NoStop}%
\bibitem [{\citenamefont {Besieris}\ \emph {et~al.}(1998)\citenamefont
  {Besieris}, \citenamefont {Abdel-Rahman}, \citenamefont {Shaarawi},\ and\
  \citenamefont {Chatzipetros}}]{Besieris98PIERS}%
  \BibitemOpen
  \bibfield  {author} {\bibinfo {author} {\bibfnamefont {I.}~\bibnamefont
  {Besieris}}, \bibinfo {author} {\bibfnamefont {M.}~\bibnamefont
  {Abdel-Rahman}}, \bibinfo {author} {\bibfnamefont {A.}~\bibnamefont
  {Shaarawi}},\ and\ \bibinfo {author} {\bibfnamefont {A.}~\bibnamefont
  {Chatzipetros}},\ }\bibfield  {title} {\bibinfo {title} {Two fundamental
  representations of localized pulse solutions to the scalar wave equation},\
  }\href@noop {} {\bibfield  {journal} {\bibinfo  {journal} {Progr. in
  Electrom. Res.}\ }\textbf {\bibinfo {volume} {19}},\ \bibinfo {pages} {1}
  (\bibinfo {year} {1998})}\BibitemShut {NoStop}%
\bibitem [{\citenamefont {Turunen}\ and\ \citenamefont
  {Friberg}(2010)}]{Turunen10PO}%
  \BibitemOpen
  \bibfield  {author} {\bibinfo {author} {\bibfnamefont {J.}~\bibnamefont
  {Turunen}}\ and\ \bibinfo {author} {\bibfnamefont {A.~T.}\ \bibnamefont
  {Friberg}},\ }\bibfield  {title} {\bibinfo {title} {Propagation-invariant
  optical fields},\ }\href@noop {} {\bibfield  {journal} {\bibinfo  {journal}
  {Prog. Opt.}\ }\textbf {\bibinfo {volume} {54}},\ \bibinfo {pages} {1}
  (\bibinfo {year} {2010})}\BibitemShut {NoStop}%
\bibitem [{\citenamefont {Kondakci}\ \emph {et~al.}(2018)\citenamefont
  {Kondakci}, \citenamefont {Yessenov}, \citenamefont {Meem}, \citenamefont
  {Reyes}, \citenamefont {Thul}, \citenamefont {Fairchild}, \citenamefont
  {Richardson}, \citenamefont {Menon},\ and\ \citenamefont
  {Abouraddy}}]{Kondakci18OE}%
  \BibitemOpen
  \bibfield  {author} {\bibinfo {author} {\bibfnamefont {H.~E.}\ \bibnamefont
  {Kondakci}}, \bibinfo {author} {\bibfnamefont {M.}~\bibnamefont {Yessenov}},
  \bibinfo {author} {\bibfnamefont {M.}~\bibnamefont {Meem}}, \bibinfo {author}
  {\bibfnamefont {D.}~\bibnamefont {Reyes}}, \bibinfo {author} {\bibfnamefont
  {D.}~\bibnamefont {Thul}}, \bibinfo {author} {\bibfnamefont {S.~R.}\
  \bibnamefont {Fairchild}}, \bibinfo {author} {\bibfnamefont {M.}~\bibnamefont
  {Richardson}}, \bibinfo {author} {\bibfnamefont {R.}~\bibnamefont {Menon}},\
  and\ \bibinfo {author} {\bibfnamefont {A.~F.}\ \bibnamefont {Abouraddy}},\
  }\bibfield  {title} {\bibinfo {title} {Synthesizing broadband
  propagation-invariant space-time wave packets using transmissive phase
  plates},\ }\href@noop {} {\bibfield  {journal} {\bibinfo  {journal} {Opt.
  Express}\ }\textbf {\bibinfo {volume} {26}},\ \bibinfo {pages} {13628}
  (\bibinfo {year} {2018})}\BibitemShut {NoStop}%
\bibitem [{\citenamefont {Yessenov}\ \emph
  {et~al.}(2020{\natexlab{b}})\citenamefont {Yessenov}, \citenamefont {Ru},
  \citenamefont {Schepler}, \citenamefont {Meem}, \citenamefont {Menon},
  \citenamefont {Vodopyanov},\ and\ \citenamefont
  {Abouraddy}}]{Yessenov20OSAC}%
  \BibitemOpen
  \bibfield  {author} {\bibinfo {author} {\bibfnamefont {M.}~\bibnamefont
  {Yessenov}}, \bibinfo {author} {\bibfnamefont {Q.}~\bibnamefont {Ru}},
  \bibinfo {author} {\bibfnamefont {K.~L.}\ \bibnamefont {Schepler}}, \bibinfo
  {author} {\bibfnamefont {M.}~\bibnamefont {Meem}}, \bibinfo {author}
  {\bibfnamefont {R.}~\bibnamefont {Menon}}, \bibinfo {author} {\bibfnamefont
  {K.~L.}\ \bibnamefont {Vodopyanov}},\ and\ \bibinfo {author} {\bibfnamefont
  {A.~F.}\ \bibnamefont {Abouraddy}},\ }\bibfield  {title} {\bibinfo {title}
  {Mid-infrared diffraction-free space-time wave packets},\ }\href@noop {}
  {\bibfield  {journal} {\bibinfo  {journal} {OSA Continuum}\ }\textbf
  {\bibinfo {volume} {3}},\ \bibinfo {pages} {420} (\bibinfo {year}
  {2020}{\natexlab{b}})}\BibitemShut {NoStop}%
\bibitem [{\citenamefont {Reivelt}\ and\ \citenamefont
  {Saari}(2002)}]{Reivelt02PRE}%
  \BibitemOpen
  \bibfield  {author} {\bibinfo {author} {\bibfnamefont {K.}~\bibnamefont
  {Reivelt}}\ and\ \bibinfo {author} {\bibfnamefont {P.}~\bibnamefont
  {Saari}},\ }\bibfield  {title} {\bibinfo {title} {Experimental demonstration
  of realizability of optical focus wave modes},\ }\href@noop {} {\bibfield
  {journal} {\bibinfo  {journal} {Phys. Rev. E}\ }\textbf {\bibinfo {volume}
  {66}},\ \bibinfo {pages} {056611} (\bibinfo {year} {2002})}\BibitemShut
  {NoStop}%
\bibitem [{\citenamefont {Guo}\ \emph {et~al.}(2021)\citenamefont {Guo},
  \citenamefont {Xiao}, \citenamefont {Orenstein},\ and\ \citenamefont
  {Fan}}]{Guo21Light}%
  \BibitemOpen
  \bibfield  {author} {\bibinfo {author} {\bibfnamefont {C.}~\bibnamefont
  {Guo}}, \bibinfo {author} {\bibfnamefont {M.}~\bibnamefont {Xiao}}, \bibinfo
  {author} {\bibfnamefont {M.}~\bibnamefont {Orenstein}},\ and\ \bibinfo
  {author} {\bibfnamefont {S.}~\bibnamefont {Fan}},\ }\bibfield  {title}
  {\bibinfo {title} {Structured {3D} linear space-time light bullets by
  nonlocal nanophotonics},\ }\href@noop {} {\bibfield  {journal} {\bibinfo
  {journal} {Light Sci. Appl.}\ }\textbf {\bibinfo {volume} {10}},\ \bibinfo
  {pages} {160} (\bibinfo {year} {2021})}\BibitemShut {NoStop}%
\bibitem [{\citenamefont {Shiri}\ \emph
  {et~al.}(2020{\natexlab{a}})\citenamefont {Shiri}, \citenamefont {Yessenov},
  \citenamefont {Webster}, \citenamefont {Schepler},\ and\ \citenamefont
  {Abouraddy}}]{Shiri20NC}%
  \BibitemOpen
  \bibfield  {author} {\bibinfo {author} {\bibfnamefont {A.}~\bibnamefont
  {Shiri}}, \bibinfo {author} {\bibfnamefont {M.}~\bibnamefont {Yessenov}},
  \bibinfo {author} {\bibfnamefont {S.}~\bibnamefont {Webster}}, \bibinfo
  {author} {\bibfnamefont {K.~L.}\ \bibnamefont {Schepler}},\ and\ \bibinfo
  {author} {\bibfnamefont {A.~F.}\ \bibnamefont {Abouraddy}},\ }\bibfield
  {title} {\bibinfo {title} {Hybrid guided space-time optical modes in
  unpatterned films},\ }\href@noop {} {\bibfield  {journal} {\bibinfo
  {journal} {Nat. Commun.}\ }\textbf {\bibinfo {volume} {11}},\ \bibinfo
  {pages} {6273} (\bibinfo {year} {2020}{\natexlab{a}})}\BibitemShut {NoStop}%
\bibitem [{\citenamefont {Guo}\ and\ \citenamefont {Fan}(2021)}]{Guo21PRR}%
  \BibitemOpen
  \bibfield  {author} {\bibinfo {author} {\bibfnamefont {C.}~\bibnamefont
  {Guo}}\ and\ \bibinfo {author} {\bibfnamefont {S.}~\bibnamefont {Fan}},\
  }\bibfield  {title} {\bibinfo {title} {Generation of guided space-time wave
  packets using multilevel indirect photonic transitions in integrated
  photonics},\ }\href@noop {} {\bibfield  {journal} {\bibinfo  {journal} {Phys.
  Rev. Research}\ }\textbf {\bibinfo {volume} {3}},\ \bibinfo {pages} {033161}
  (\bibinfo {year} {2021})}\BibitemShut {NoStop}%
\bibitem [{\citenamefont {Shiri}\ \emph
  {et~al.}(2020{\natexlab{b}})\citenamefont {Shiri}, \citenamefont {Yessenov},
  \citenamefont {Aravindakshan},\ and\ \citenamefont {Abouraddy}}]{Shiri20OL}%
  \BibitemOpen
  \bibfield  {author} {\bibinfo {author} {\bibfnamefont {A.}~\bibnamefont
  {Shiri}}, \bibinfo {author} {\bibfnamefont {M.}~\bibnamefont {Yessenov}},
  \bibinfo {author} {\bibfnamefont {R.}~\bibnamefont {Aravindakshan}},\ and\
  \bibinfo {author} {\bibfnamefont {A.~F.}\ \bibnamefont {Abouraddy}},\
  }\bibfield  {title} {\bibinfo {title} {Omni-resonant space-time wave
  packets},\ }\href@noop {} {\bibfield  {journal} {\bibinfo  {journal} {Opt.
  Lett.}\ }\textbf {\bibinfo {volume} {45}},\ \bibinfo {pages} {1774} (\bibinfo
  {year} {2020}{\natexlab{b}})}\BibitemShut {NoStop}%
\bibitem [{\citenamefont {Shiri}\ \emph
  {et~al.}(2020{\natexlab{c}})\citenamefont {Shiri}, \citenamefont {Schepler},\
  and\ \citenamefont {Abouraddy}}]{Shiri20APLP}%
  \BibitemOpen
  \bibfield  {author} {\bibinfo {author} {\bibfnamefont {A.}~\bibnamefont
  {Shiri}}, \bibinfo {author} {\bibfnamefont {K.~L.}\ \bibnamefont
  {Schepler}},\ and\ \bibinfo {author} {\bibfnamefont {A.~F.}\ \bibnamefont
  {Abouraddy}},\ }\bibfield  {title} {\bibinfo {title} {Programmable
  omni-resonance using space-time fields},\ }\href@noop {} {\bibfield
  {journal} {\bibinfo  {journal} {APL Photon.}\ }\textbf {\bibinfo {volume}
  {5}},\ \bibinfo {pages} {106107} (\bibinfo {year}
  {2020}{\natexlab{c}})}\BibitemShut {NoStop}%
\end{thebibliography}%

\end{document}